\documentclass[a4paper,11pt]{article}
\usepackage{jcappub}
\usepackage{lineno}
\usepackage{amsmath}	
\usepackage{bm}
\usepackage{color}

\def\vk{\bm{k}}
\def\vq{\bm{q}}

\newcommand{\be}{\begin{equation}}
\newcommand{\ee}{\end{equation}}
\newcommand{\bea}{\begin{eqnarray}}
\newcommand{\eea}{\end{eqnarray}}
\def\ba#1\ea{\begin{align}#1\end{align}}

\def\({\left(}
\def\){\right)}

\def\[{\left[}
\def\]{\right]}

\newcommand{\refeq}[1]{Eq.~(\ref{eq:#1})}

\newcommand{\reffig}[1]{Fig.~\ref{fig:#1}}
\newcommand{\refsec}[1]{Sec.~\ref{sec:#1}}

\newcommand{\reftab}[1]{Tab.~\ref{tab:#1}}
\newcommand{\vs}{\nonumber\\}

\definecolor{RedWine}{rgb}{0.743,0,0}
\definecolor{GrassGreen}{rgb}{0.125,0.75,0.125}

\arxivnumber{2204.00668}
\title{Spherical Bispectrum: A Novel Visualization Scheme For Facilitating Comparisons}

\author[a,b]{Joseph Tomlinson}
\author[a,b,c]{Donghui Jeong}
\affiliation[a]{Department of Astronomy and Astrophysics, The Pennsylvania State University, University Park, PA 16802, USA}
\affiliation[b]{Institute for Gravitation and the Cosmos, The Pennsylvania State University, University Park, PA 16802, USA}
\affiliation[c]{School of Physics, Korea Institute for Advanced Study, 85 Hoegiro, Dongdaemun-gu, Seoul, 02455, Republic of Korea}

\emailAdd{jxt732@psu.edu}

\abstract{Recent developments of Perturbation Theory (PT), specifically the Effective Field Theory of Large Scale Structure (EFTofLSS) and its equivalents, have proven powerful in analyzing galaxy clustering statistics such as the galaxy power spectrum and bispectrum. To further this pursuit, we have devised a novel spherical-bispectrum visualization scheme that collapses configuration dependencies to highlight the scale dependence of the bispectrum. The resulting one-dimensional curves facilitate the comparison between different bispectra, for example, from simulation and PT calculation. Using the new scheme, we present a quantitative analysis of the accuracy of PT modeling by comparing PT's analytical prediction to the result from a suite of Quijote simulations. Specifically, we determine $k_{\rm NL}$,  the wavenunmber below which the analytical prediction matches well with the N-body result by inspecting 
both leading order (LO) and next-to-leading order (NLO) power spectrum and bispectrum at redshifts $z=0$, $0.5$, $1$, $2$, $3$. We also quantify the binning effect in Fourier space and show that an appropriate correction must be applied to the analytic predictions in order to compare them with the discrete Fourier transform results obtained from N-body-simulation or real data.}

\begin{document}
\maketitle
\flushbottom

\section{Introduction} \label{sec:intro}

Upcoming large-scale structure (LSS) surveys \citep{dore2015cosmology,hill2008hobbyeberly,laureijs2011euclid,levi2013desi,lsstdarkenergysciencecollaboration2012large,maartens2015cosmology,spergel2015widefield,Takada_2014} will allow us to probe the clustering of galaxies on ever larger scales while measuring them on existing scales with high precision. This increase in precision on small-scale galaxy clustering measurements calls for more accurate models that have well characterized systematics. There has been significant development of the theoretical models in recent years, from fast methods for computing next-to-leading and higher-order corrections \citep{Schmittfull_2016,McEwen_2016,slepian2018decoupling,Schmittfull_2016two,Simonovi__2018}, to developments of the galaxy bias models to bridge theory and observations \citep{Scoccimarro_2004,Taruya_2010,Zheng_2010,Hirata_2009,Blazek_2015, Eggemeier_2019}. See \cite{Desjacques_2018review} for a review, and \cite{Desjacques_2018,Tomlinson_2020,Ivanov_2020} for models combining both.

While most of this work has been done on the power spectrum, this increase in precision will also allow for higher-order statistics to be used, typically the three-point function or the bispectrum. Some significant progress has been made in modelling the bispectrum, the Fourier transform of the three-point function, in a similar way to the power spectrum \citep{Eggemeier_2019,Smith_2008,Rampf_2012,Assassi_2014,Bernardeau_2012,Lazanu_2016,Angulo_2015,eggemeier2021testing,McCullagh/Jeong/Szalay:2015}. 
Using the bispectrum is important for properly breaking degeneracies that are present in the two-point function or power spectrum. The use of the bispectrum has been shown to increase constraints on various cosmological parameters of importance, dark energy \citep{Sefusatti_2006,Song_2015,Byun_2017}, primordial non-Gaussianity \citep{Sefusatti_2012,Tellarini_2016,karagiannis2020probing,dizgah2020primordial}, and neutrino mass \citep{Chudaykin_2019,Hahn_2020,hahn2020constraining,kamalinejad2020nondegenerate}. The bispectrum has been previously applied to data \citep{Gil_Mar_n_2016,Slepian:2015hca,Slepian_2017,Cabass/etal:2021}, but only very recently has this been done on anything but large scales where it is most well modeled. Though recently \cite{eggemeier2021testing} has shown results past the linear regime.

To exploit the full potential of the dataset, these models often go beyond leading order (LO) to the next-to-leading order (NLO) in perturbation theory, allowing for accurate modelling of smaller-scale clustering signatures; for a review see \cite{Bernardeau_2002}. The exact scale at which NLO perturbation theory is required is extremely important not only for accurately estimating the information gained by including NLO corrections, but also for properly keeping theoretical systematics under control. This has been studied to some extent for the power spectrum \citep{Jeong_2006,Nishimichi_2009,Taruya_2009,Osato_2019} and bispectrum \citep{Lazanu_2016,Steele_2021, Lazanu_2018, Baldauf_2021}, but extensive measurements for a variety of redshifts are required, since the range of validity is sensitive to the redshift. This is especially true for the bispectrum. With accurate knowledge of the difference between the applicable scales of the LO and NLO models, along with knowledge of covariance matrices (e.g. \cite{Villaescusa_Navarro_2020}) to measure mode coupling at small scales which erases some of the information gained from the extra modes at those scales, one could then accurately predict the constraining power gained from using a NLO model over a LO model.

This is the main purpose of our new visualization scheme introduced in this paper. Since the bispectrum is a function of three variables, direct comparisons between two different models is more challenging compared to the one variable power spectrum. We alleviate this problem by introducing a transformation scheme that takes all the scale information and collapses it into a single variable, while putting shape information into other variables. This is done by doing a weighted binning of different modes that have similar scales which should have minimal information loss. We title this transformation/visualization scheme as the "Spherical Bispectrum" since if we treat the original bispectrum variables as equivalent to a standard 3D Cartesian coordinate system then our transformed coordinates are almost exactly the corresponding spherical coordinates. Since all of the scale information is contained within a single variable this allows us to compute deviations between two bispectra in an identical way to the power spectrum.

In this paper we shall measure $k_{\rm NL}$, the maximum wavenumber below which perturbation theory provides an accurate modeling of the nonlinear density field, by comparing the matter power spectrum and bispectrum obtained from a suite of Quijote N-body simulation \cite{Villaescusa_Navarro_2020}. We have defined $k_{\rm NL}$ by requiring the cumulative deviation of perturbation theory prediction to stay below 1~\% and 2~\% for the power spectrum, and 2~\% and 5~\% for the bispectrum. We repeat the analysis for the LO prediction and NLO prediction to obtain $k_{\rm NL}$ for each case. We also study the triangular-configuration dependencies of $k_{\rm NL}$ for the nonlinear bispectrum.

This paper is organized as follows. We start in \refsec{data} with providing details on the data we used from the Quijote N-body simulations and the basics of the theory we used to match them, including binning effects. We then present our new method for visualizing the bispectrum, which we denote the spherical bispectrum, in \refsec{bksph}. Finally our results for the range of validity for both the power spectrum and the bispectrum, including configurations, at all redshifts is included in \refsec{scale}. We conclude in \refsec{conclusion}.

Throughout the paper, we use the following conventions and shorthand notations. Our definition of the bispectrum is
\be
\left<
\delta(\vk_1)\delta(\vk_2)\delta(\vk_3)
\right>
=
(2\pi)^3B(\vk_1,\vk_2,\vk_3)
\delta^D(\vk_{123})\,.
\ee
Here, $\delta^D$ is the Dirac-delta operator, and we use the shorthanded notation of $\vk_{1\cdots n}\equiv \vk_1+\cdots\vk_n$. We denote the amplitude of a vector $\vk_i$ as $k_i$.

\section{Data \& Theory}\label{sec:data}

\subsection{Quijote Simulations}

For our N-body data set we use a subset of the Quijote simulations \citep{Villaescusa_Navarro_2020}, a set of 43,100 N-body simulations with a wide variety of cosmological parameters and results available at five different redshifts ($z=0$, $0.5$, $1$, $2$, and $3$). For this study, we use the high-resolution, fiducial-cosmology sample, a set of 100 simulations run with 8 times as many dark matter particles.
In detail, this sample of 100 simulations each has $1024^3$ dark matter particles distributed over a volume of 1 Gpc$^3$/$h^3$, giving a fundamental wavenumber $k_f \approx 0.0063$ $h$/Mpc, with a cosmology in agreement with \cite{Planck2018}. For more details about the simulations themselves, see \cite{Villaescusa_Navarro_2020}. 

We choose the high resolution simulations in particular because the higher resolution gives better accuracy at small scales, where we need it to determine the range of validity, at the cost of large scale error bars due to only having 100 simulations. We find this tradeoff to be worthwhile since it is already well known that LO and NLO perturbation theory are consistent with the nonlinear density field on large scales, so large error bars there do not impact the results of our analysis. In this work, we have devised a criteria for defining $k_{\rm NL}$ without being affected by the deviation due to large-scale cosmic  variance. From these simulations we use the precomputed power spectra \citep{Villaescusa_Navarro_2020} and then use the full dark-matter particle distribution to measure the bispectrum, all of which exists for five redshifts. The reason we remeasure the bispectrum is because the precomputed bispectra are not minimally binned, with $\Delta k = 2k_f$, so we remeasure them with minimal binning, $\Delta k = k_f$ to ensure the most accurate representation of the underlying data we can get and to minimize binning effects.

An important part of measuring the bispectrum is distributing the N-body particles onto a grid. For our grid decomposition we use \textit{nbodykit}\citep{Hand_2018} with a custom \textbf{Julia} wrapper. We also follow the recommendations in \cite{Sefusatti_2016} and use interlaced grids along with the third-order triangular-shaped clouds (TSC) method to dramatically reduce aliasing effects. 
For measuring the bispectrum from the decomposed grid we use the fast Scoccimarro estimator \citep{Scoccimarro_2015,Sefusatti_2016,Tomlinson_2019}, specifically we use the code presented in \cite{Tomlinson_2019} with some efficiency improvements. We measure the bispectrum up to $k_{\rm NL} = 80k_f \approx 0.503$ $h$/Mpc to ensure we have sufficiently small enough scales to properly capture the deviation from perturbation theory at high redshift.

\subsection{Eulerian Perturbation Theory} \label{sec:PTmodel}

As for the theory model, we use the standard Eulerian perturbation theory(SPT) model, see \cite{Bernardeau_2002} for a review. In brief, SPT factors the full nonlinear matter density into different parts 
\be
\delta(\vk) = \delta_L(\vk) + \delta^{(2)}(\vk) + \delta^{(3)}(\vk) +\cdots\,,
\ee
where $\delta^{(n)}(\vk)$ stands for the $n$-th order density contrast proportional to the $n$ linear quantities such as density field, velocity field, and tidal field.
We can then use Wick's theorem to find the leading order (LO) and the next-to-leading order (NLO) contribution to the power spectrum straightforwardly
\ba
\left< \delta(\vk) \delta(\vk') \right> &= \left< \delta_L(\vk) \delta_L(\vk') \right> 
+ 2\left< \delta_L(\vk)\delta^{(3)}(\vk') \right> + \left< \delta^{(2)}(\vk)\delta^{(2)}(\vk') \right>
\ea
We write each term in the above equation as
\be
P_{\rm LO+NLO}(k) = P_L(k) + 2P_{13}(k) + P_{22}(k)\,,
\label{eq:pkNLO}
\ee
where the leading order power spectrum is simply the linear power spectrum $P_{\rm LO}(k)=P_{L}(k)$, and the NLO terms consist of
\ba
P_{22}(k) &= 2\int\frac{{\rm d}^3 \vq}{(2\pi)^3}F_2(\vq, \vk - \vq)^2P_L(q)P_L(|\vk - \vq|)
\vs
P_{13}(k) &= 3P_L(k)\int\frac{{\rm d}^3 \vq}{(2\pi)^3} F_3(\vq, -\vq, \vk)P_L(q) \,.
\ea
Here, $F_n(\vq_1,\cdots,\vq_n)$ is the $n$-th order integration kernel, see \cite{Bernardeau_2002}. For the form suitable for a faster implementation of these integrals using the FFTlog-based method, see \cite{Schmittfull_2016,McEwen_2016}.

Since the initial density field is close to Gaussian \citep{WMAP2013,Planckfnl2020}, we assume that the linear-order density field follows Gaussian statistics, and there is no linear expression for the bispectrum. The leading-order bispectrum expression is 
\ba \label{eq:blo}
B_{{\rm LO}}(k_1,k_2,k_3) =
B_{211}(k_1,k_2,k_3) 
=
\, 2F_2(\vk_1,\vk_2)P_L(k_1)P_L(k_2) + (2~{\rm cyclic})\,.
\ea
Like the NLO power spectrum calculation, we can use Wick's theorem to find the NLO contribution to the bispectrum
\ba
&B_{\rm LO + NLO}(k_1,k_2,k_3) 
= B_{\rm LO}(k_1,k_2,k_3) 
+ B_{222}(k_1,k_2,k_3) 
\vs &
+ B_{411}(k_1,k_2,k_3) + B_{123}^{I}(k_1,k_2,k_3) + B_{123}^{II}(k_1,k_2,k_3)\,,
\label{eq:BkNLO}
\ea
where
\ba
 B_{222}(k_1,k_2,k_3)
&=\, 8 \int \frac{{\rm d}^3 \vq}{(2\pi)^3} F_2(\vq, \vk_1 - \vq)
   F_2(-\vq,\vk_2+\vq)
   F_2(\vq-\vk_1,-\vk_2-\vq)
\vs &\hspace{2cm} \times
P_L(q)P_L(|\vk_1-\vq|)P_L(|\vk_2+\vq|)\,,
\\
 B_{411}(k_1,k_2,k_3)
&=\, 12 P_L(k_1)P_L(k_2)\int\frac{{\rm d}^3\vq}{(2\pi)^3}
F_4(\vq,-\vq,-\vk_1,-\vk_2)P_L(q) 
+ ({\rm 2 \,cyclic})\,,
\\
 B_{123}^{I}(k_1,k_2,k_3)
&=\, 6 P_L(k_1)\int\frac{{\rm d}^3 \vq}{(2\pi)^3} F_2(\vq,\vk_2-\vq)
F_3(-\vq,\vq-\vk_2,-\vk_1)
\vs &\hspace{3.2cm}\times 
P_L(q)P_L(|\vk_2-\vq|) + ({\rm 5 \,cyclic})\,,
\\
B_{123}^{II}(k_1,k_2,k_3)
&=\, 6P_L(k_1)P_L(k_2)F_2(\vk_1,\vk_2)
\int\frac{{\rm d}^3 \vq}{(2\pi)^3} F_3(\vq, -\vq, \vk_2)P_L(q) + (\rm{5 \,cyclic})\,.
\ea

A fast way to compute these integrals exists \citep{IASMatrixFFT2018}, but not one using the standard formalism and has not been extended to redshift-space or biased tracers.

These perturbation theory expressions are what we use throughout this work. For the linear power spectrum we take the one provided with the Quijote simulations generated using CAMB \citep{Lewis_2000}. To calculate the NLO power spectrum model we use an adaptation of the methods described in \cite{Schmittfull_2016}. The LO bispectrum model is generated using the provided linear power spectrum and \refeq{blo}. We generate our NLO bispectrum model through direct integration.

\subsection{Binned Theory Models}

\begin{figure*}
    \centering
    \includegraphics[width=0.485\textwidth]{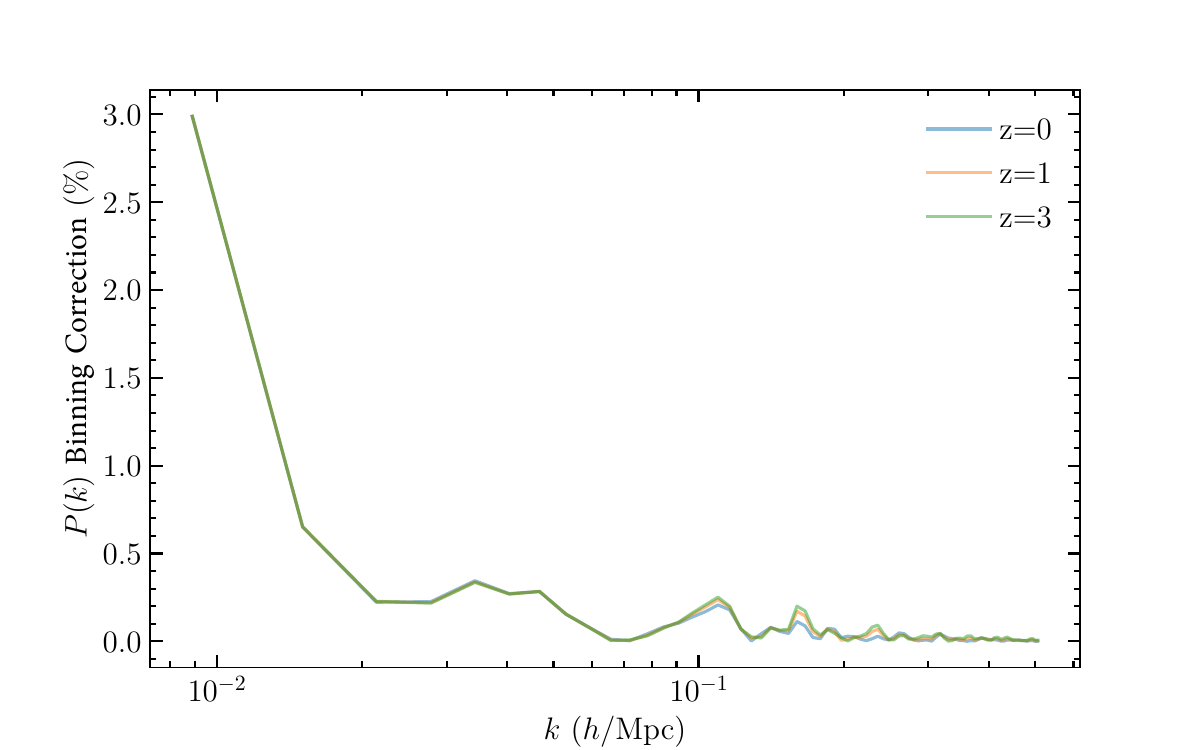}
    \includegraphics[width=0.485\textwidth]{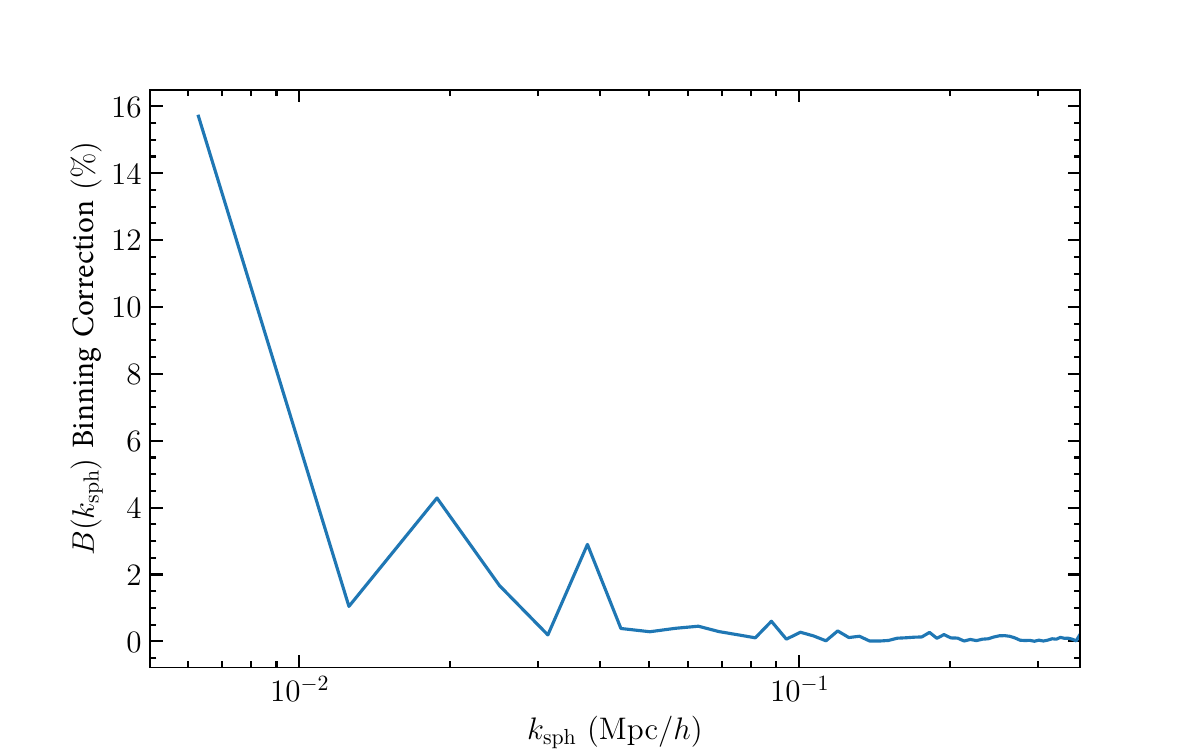}
    \caption{({\it Left}): The binning correction on the LO + NLO power spectrum. The binning correction is plotted for three plotting redshifts ($z=0$, $1$, $3$) to emphasize the contributions originating from the NLO part of the model, since only the NLO part has any redshift evolution. We see that binning correction matters only for very large scales, as it stays below 1\% in all bins except the largest scale bin. 
    ({\it Right}): The binning correction on the LO + NLO bispectrum, as a function of $k_{\rm sph}$ that we define later in \refsec{bksph}. The visible binning correction is independent of redshift, signifying that the NLO contribution is much smaller than the LO contribution, so we only plot one redshift at $z=0$. We see significantly larger errors than the case of the power spectrum with the largest bin reaching an error around 16\,\%. }
    \label{fig:binning}
\end{figure*}

Fourier-space clustering statistics measured from a density field defined at grid points are inherently binned with the fundamental frequency $k_F=2\pi/L$. On the other hand, analytic theory models, like those in \refsec{PTmodel}, can be calculated at each wavenumber without binning. This causes a discrepancy between the two even if the underlying statistic is identical, so a correction is essential when comparing analytic theory models and binned measurements. We refer to this as a binning correction. Note that theory models that are inherently gridded, such as the result from GridSPT \citep{Taruya_2018}, do not suffer from this problem but instead they are statistical and so require many realizations to reduce statistical errors. 

To compute the binning correction we need to bin the analytic models the exact same way that we bin the statistical measures from simulations or real data. This is typically done by using {\it an estimator}, but, instead of taking data as input, the theory-model calculation is used as input. We then quantify the binning correction as the difference between the binned and unbinned theory relative to the unbinned theory. In this section, we shall measure the binning correction for the power spectrum and bispectrum.

For the power spectrum, the binning correction is typically very small except on scales near the fundamental wavenumber. To bin the theoretical power spectrum, we first generate a three-dimensional grid in Fourier space and assign the LO and NLO power spectrum calculated by using \refeq{pkNLO} to each grid point.  We then measure the binned theory model for both LO and NLO power spectrum 
the same binning scheme as the direct measurement:
\be
P_{\rm bin}(k_i) = \frac{1}{N_k}\sum\limits_{|\vk-k_i|\leq \frac{1}{2}k_F} P(\vk) \,.
\ee
Note that this is just the traditional direct power spectrum estimator but with $|\delta(\vk)|^2$ replaced with $P(\vk)$. 
We show the binning-correction error for the LO+NLO power spectrum in the left pane of \reffig{binning}. We have plotted the errors for our three plotting redshifts ($z=0$, $1$, $3$) to emphasize the contributions coming from the NLO part, those that evolve with redshift, while the contributions that do not evolve with redshift are from the LO part, the linear power spectrum. We see that the binning effect contributes overwhelmingly more to the LO model, likely because it is only significant for the very largest scales. Our results are broadly consistent with the folklore that the binning correction only matters on very large scales, with only the very largest bin having an error larger than 1\%. However, to use the large-scale bins for precision work, for example for measuring the local-type primordial non-Gaussianity \citep{Slosar_2008,Hamaus2011,Giannantonio2014,Agarwal_2014,Barreira2020,Morad2021, Rezaie_2021}, an accurate modeling of the binning correction is necessary.

For the bispectrum the binning effect is significantly more important than the power spectrum. \cite{Bernardeau_2012} and more recently \cite{eggemeier2021testing} have reported significant binning effects between a few and up to 10 percent and its triangular-configuration dependencies. The binning effect on the bispectrum is computationally expensive to study, because we must sum over all theory contribution on the grid:
\ba
B_{\rm bin}(k_1,k_2,k_3) =& 
\frac{1}{N_{123}} 
\sum\limits_{|\vk_i - k_1|\leq \frac{1}{2}k_F}
\sum\limits_{|\vk_j - k_2|\leq \frac{1}{2}k_F}
\sum\limits_{|\vk_k - k_3|\leq \frac{1}{2}k_F}
B(\vk_i,\vk_j,\vk_k)\,.
\label{eq:Bk_binning}
\ea
For example, for the bispectra considered in this work, it takes around a week for each of LO and NLO bispectrum models to get the binned form. 

The right panel of \reffig{binning} shows the binning correction for the NLO bispectrum, as a function of the $k_{\rm sph}$ that we define in \refsec{bksph}. Unlike the power spectrum we do not plot this for multiple redshifts, since the contribution from the NLO part of the bispectrum, and hence the redshift evolution, is indistinguishably small on this plot. This means that nearly all of the binning effect, at least in $k_{\rm sph}$-space, is due to the LO part. When plotted in $k_{\rm sph}$-space, we see the bispectrum display similar behavior to the power spectrum, with significant binning errors on very large scale modes, reaching a maximum error of 16\% in the largest bin.

Although we use the computationally slow exact binning in this work, as we only needed to compute it a few times, the long computation time of binning the bispectrum is a clear obstacle for precision bispectrum analysis given how significant the effect can be. A few alternative techniques have been developed. \cite{Sefusatti_2010} have developed a method by averaging over the wavenumbers in a bin to generate an effective wavenumber
\ba
k_{{\rm eff},i}(k_1,k_2,k_3)=& 
\frac{1}{N_{123}} 
\sum\limits_{|\vk_i - k_1|\leq \frac{1}{2}k_F}
\sum\limits_{|\vk_j - k_2|\leq \frac{1}{2}k_F}
\sum\limits_{|\vk_k - k_3|\leq \frac{1}{2}k_F}
\vk_i\,,
\ea
and evaluating the theoretical-model bispectrum at that effective wavenumber. Following \cite{Yankelevich2018}, \cite{Oddo_2020} improves the method by sorting the wavenumbers, achieving a less than 5\% binning error. Recently, \cite{eggemeier2021testing} has introduced a scheme which uses Delaunay interpolation and constructs tetrahedra to reduce the computation time, achieving around 1\% accuracy when using full Delaunay binning.

Finally, a comment is in order for the interpolation scheme that we use to compute the binned power spectrum and bispectrum. When computing the power spectrum and the NLO bispectrum at the grid points, we use cubic-spline interpolation to alleviate needing to compute the model at each grid point. For the power spectrum, which is a very smooth function, this is unlikely to cause any significant effects. For the bispectrum, however, our testing shows this can have a significant effect. For example, we find $\sim$ 10\% error on binning correction at the largest scales when computing \refeq{Bk_binning} with the interpolated LO bispectrum instead of the exact LO bispectrum. Note, however, that we only use interpolation for the NLO-bispectrum binning correction, which is not very significant on any scales. Therefore, the interpolation should not affect our results.

\section{The Spherical Bispectrum} \label{sec:bksph}

\begin{figure*}
    \centering
    \includegraphics[width=0.495\textwidth]{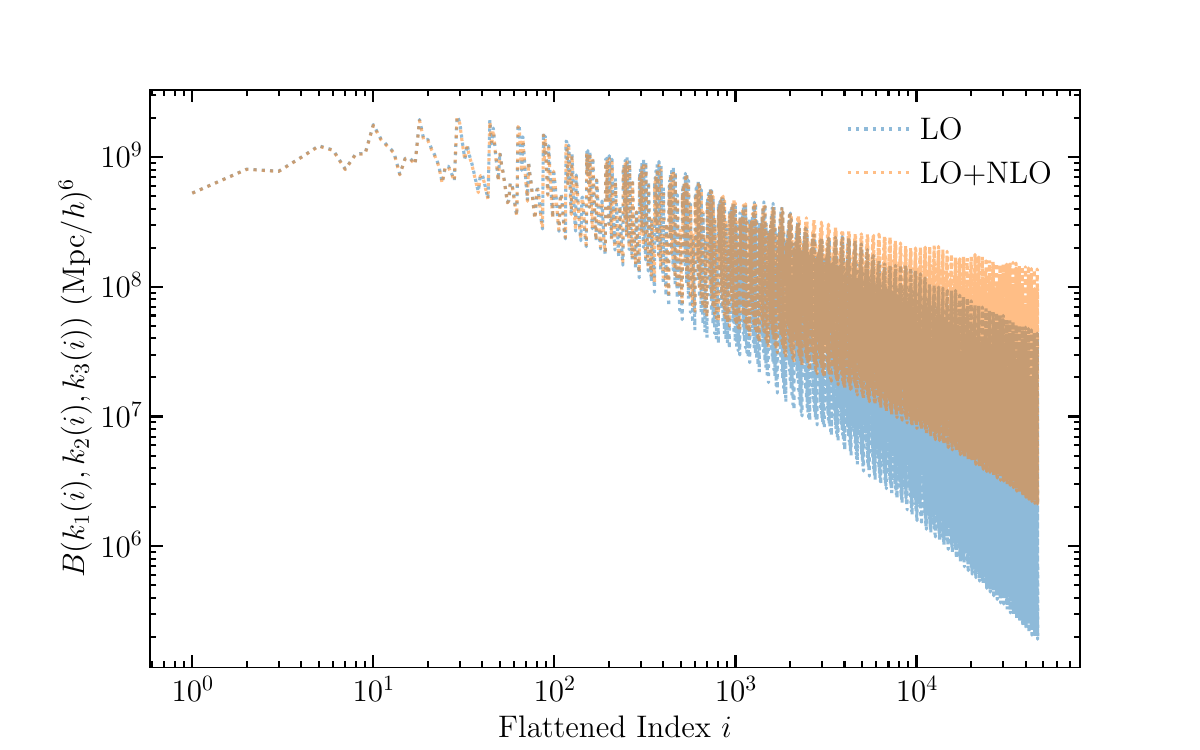}
    \includegraphics[width=0.495\textwidth]{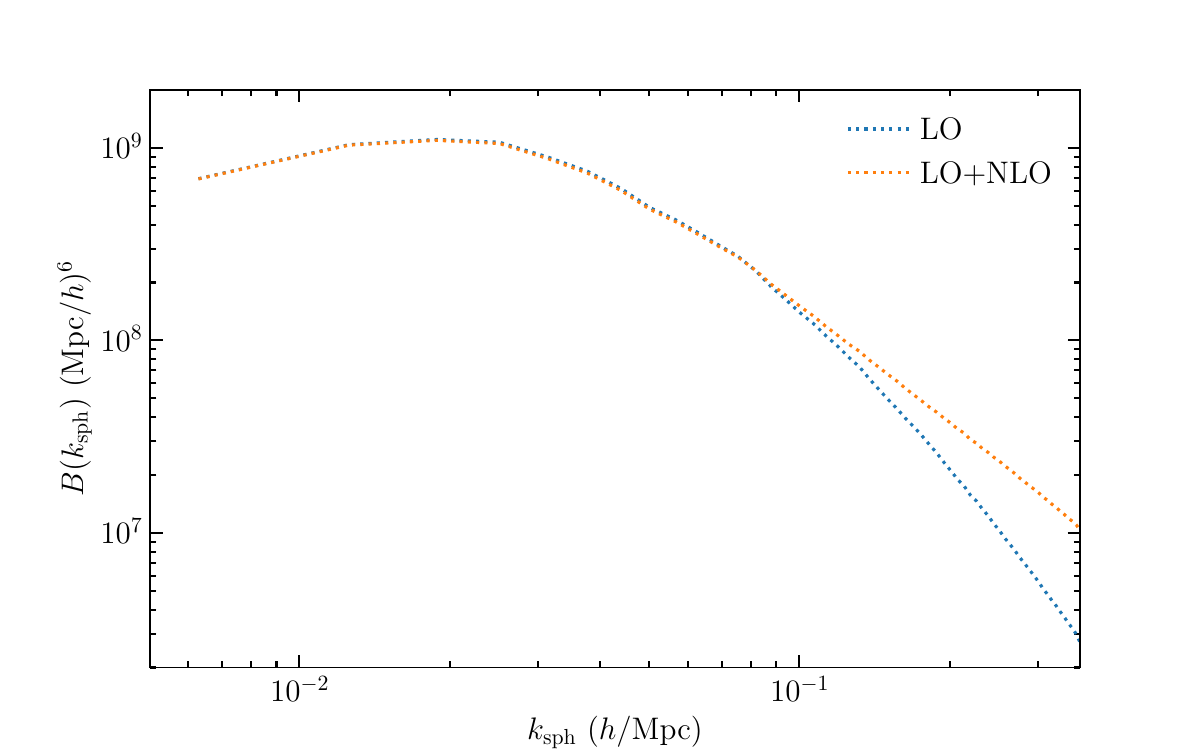}
    \caption{({\it Left}): The LO and NLO bispectrum models plotted in the usual flattened index space. The flattened index is generated from creating an ordered set of tuples ($k_1,k_2,k_3$) with $k_1\geq k_2\geq k_3$ and incrementing in row-major order, that is incrementing $k_3$ first. Binned with bin size $k_f$
     ({\it Right}): The same two models plotted in spherical bispectrum space. It is both much easier to see the differences between the models and also more clear at exactly what scale they begin to diverge. Both plots are at $z=0$.
     }
    \label{fig:flatindex}
\end{figure*}

The Bispectrum is a function of three semi-independent variables, $k_1$, $k_2$, and $k_3$, connected by the triangle condition: $k_a+k_b\le k_c$ for all combination of $a\neq b\neq c$. While this does not hamper the bispectrum's ability as a cosmological probe, it is desirable to have a visualization tool showing the full scale- and configuration- dependence and facilitating the comparison among different models.

In literature, the most common method of visualizing the bispectrum is by flattening the three input wavenumbers into a contiguous one-dimensional flattened index each of which maps onto the triplet ($k_1$, $k_2$, $k_3$). The flattened index plot is not smooth when $k_1$ and $k_2$ values vary so the plot features are dominated by indexing effects rather than the nonlinearities. Also, it is non-trivial to assign a numeric scale to a specific bispectrum element, since there are three separate scales that contribute towards it. As a result, it is difficult to assign a single $k_{\rm NL}$ by examining the flattened index plot.

Other methods, such as heatmap triangle plots \citep{Jeong/Komatsu:2009,McCullagh/Jeong/Szalay:2015}, also exist although they can only be done by holding one of the wavenumbers constant. Thus, while highlighting the configuration-dependence of the bispectrum, the latter visualization methods are not convenient for direct comparison among different bispectra, for example, from theory calculation and simulation measurement.

To facilitate the comparison between different models with each other and against data visually, here we introduce the {\it spherical bispectrum}, which is essentially a coordinate transformation from the wavenumber triplet ($k_1$, $k_2$, $k_3$) to the spherical coordinate system ($k_{\rm sph}$, $\theta_{\rm sph}$, $\phi_{\rm sph}$). That is, by treating the wavenumbers of the bispectrum as equivalent to a Cartesian coordinate system (i.e. $k_1 = k_x$, $k_2=k_y$, $k_3=k_z$), we can then convert this three-dimensional coordinate system into spherical coordinates like so
\ba
k_{\rm sph} =& \sqrt{k_1^2 + k_2^2 + k_3^2} / \sqrt{3}
\vs
\phi_{\rm sph} =& \arctan\(k_2/k_1\)
\vs
\theta_{\rm sph} =& 
\arctan\(\sqrt{k_1^2+k_2^2}/k_3\)\,.
\ea

We then bin the bispectrum results $B(k_{\rm sph},\phi_{\rm sph},\theta_{\rm sph})$ based on $k_{\rm sph}$ and compute a weighted average off all the modes that fall into the same bin, weighted by the number of triangles that contribute to each mode. Formally, we may write the spherical bispectrum as
\be
B_{\rm sph}(k_{\rm sph}) = 
\frac{\sum\limits_{|q_{\rm sph} - k_{\rm sph}| < \Delta/2} B(q_1,q_2,q_3) N_{\rm tri}(q_1,q_2,q_3)}{\sum\limits_{|q_{\rm sph} - k_{\rm sph}| < \Delta/2} N_{\rm tri}(q_1,q_2,q_3)}\,,
\ee
with the bin size $\Delta$. The spherical bispectrum allows us to get a good measure of the bispectrum at a given scale, which is also easily plot-able in a standard form identical to the power spectrum. The $k_{\rm sph}$ can be thought of as a geometrically motivated way of defining a single {\it effective wavenumber} for each wavenumber triplet $(k_1,k_2,k_3)$.

In \reffig{flatindex}, we plot two theory curves, the leading order (LO) and next-to-leading order (NLO) bispectrum in two different visualization schemes: the traditional flattened index style (Left) and the new spherical bispectrum style (Right).
As shown in \reffig{flatindex}, we see significantly enhanced clarity with regards to where the two models deviate when looking at them in $k_{\rm sph}$-space compared to flattened-index space. 

Note that if we only use the radial binning, the spherical bispectrum does lose some information compared to using the full bispectrum.

We can differentiate, however, the bispectrum's configuration dependence by taking advantage of the angular information in the spherical coordinates. In \refsec{config}, we use six triangular configurations (five used in the previous bispectrum analysis along with a new one), which are differentiated based on angular cuts in the spherical coordinates, giving us easy visual access to different parts of the angular domain.

\subsection{Spherical Triangle Configurations} \label{sec:config}

\begin{figure*}
    \centering
    \includegraphics[width=0.495\textwidth]{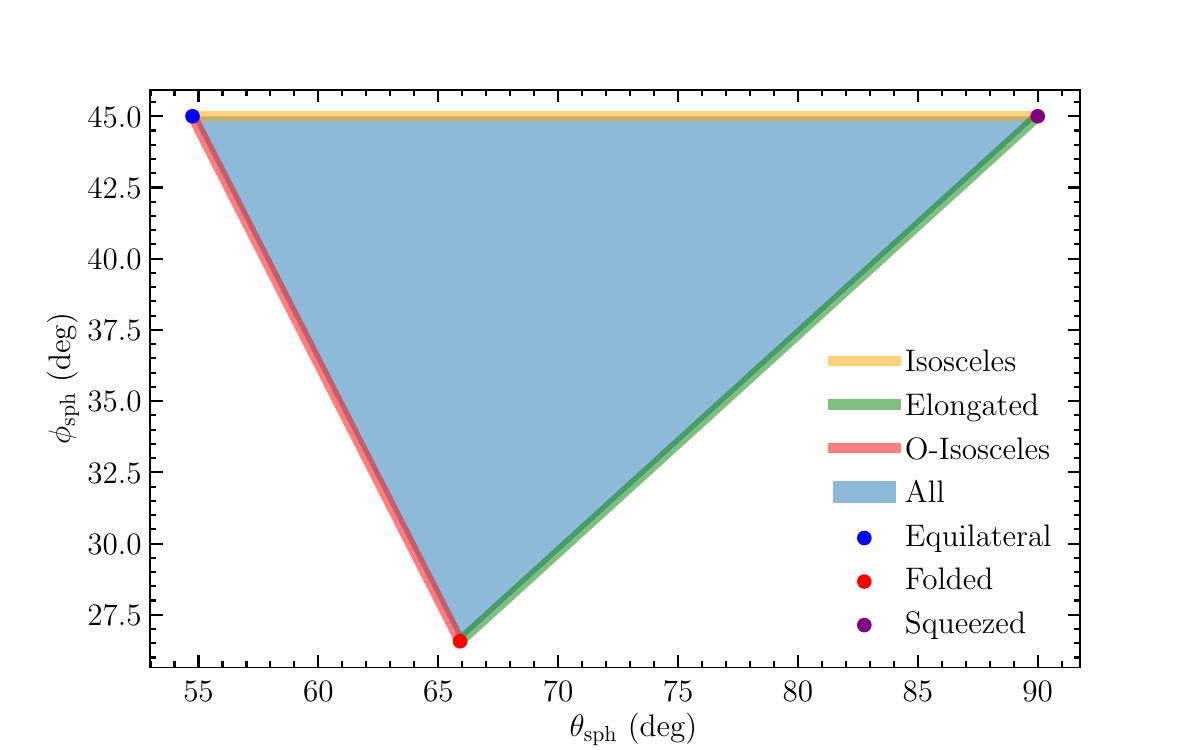}
    \includegraphics[width=0.495\textwidth]{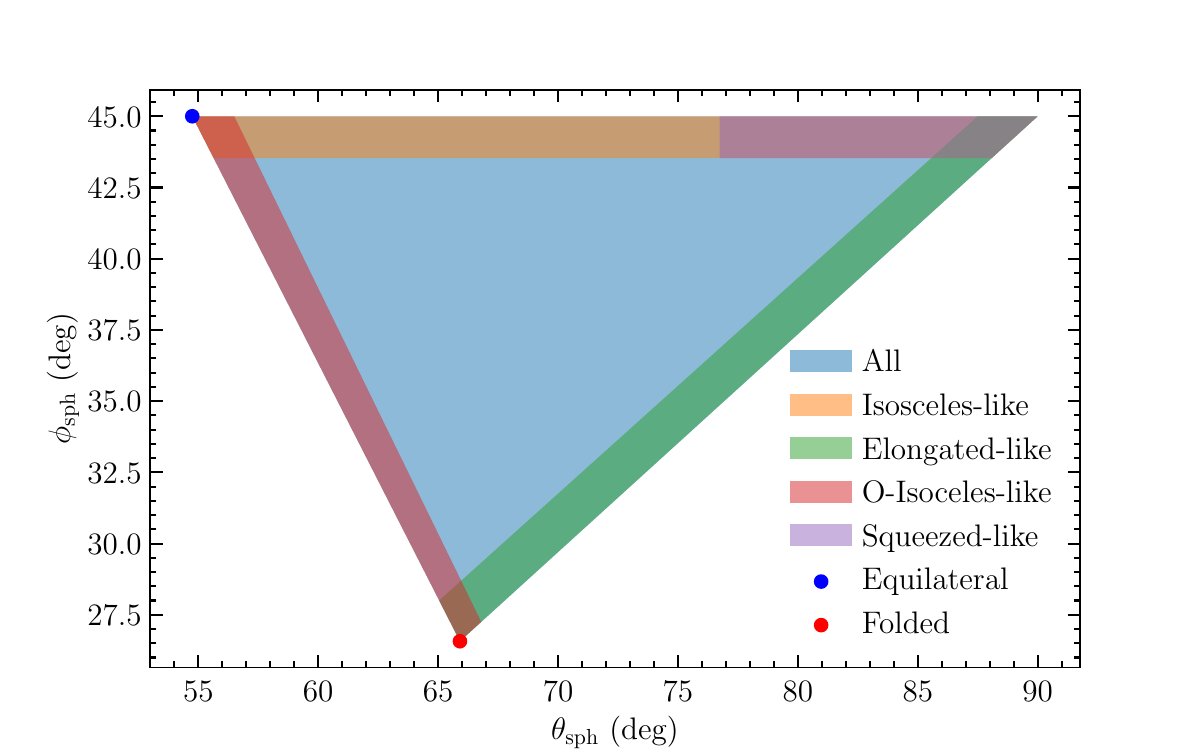}
    \caption{On the left is the formal definintions of each config in $\theta_{\rm sph}-\phi_{\rm sph}$ space. On the right is our expanded definitions for ``-like" configs we use in this work. Here we have relaxed equality conditions by 5\% and consider any mode with $k_1/k_3 > 3$ to be squeezed.}
    \label{fig:angle}
\end{figure*}

Probing specific range of angles in spherical coordinates allows us to look at the bispectrum's specific triangular configurations and to regain the information that is lost in the radial binning procedure. These configuration dependencies have been studied as ways to probe different physical mechanism of generating the bispectrum because the bispectrum at each configuration responds differently for different physical origins such as early Universe physics, non-linear growth of structure, and nonlinear bias \citep{Desjacques_2018review}.

There are six primary configurations we study here, and they are formally defined as follows:
\begin{itemize}
\item Equilateral:
$
k_1=k_2=k_3
$
\item Isosceles:
$
k_1=k_2>k_3
$
\item Folded:
$
k_1 = 2k_2 = 2k_3
$
\item Elongated:
$
k_1=k_2+k_3
$
\item Squeezed:
$
k_1=k_2 \gg k_3
$
\item Obtuse-Isosceles (O-Isosceles): 
$
k_1 > k_2 = k_3
$
\end{itemize}
However, some of these configurations contain too few modes per bin and, even with reasonable bin sizes, using their strict definition makes them very stochastic. The problem is particularly severe for the Squeezed where the equality holds up for only limited number of configurations. 

We alleviate this problem, and also include more information, by expanding their definitions while still only including modes that capture the same type of mode ratios. When used in this way we refer to them as ``-like" configurations, for the remainder of this section, to differentiate from the formal definitions given above. In \reffig{angle}, we show the different configurations in $\theta_{\rm sph}-\phi_{\rm sph}$ space, both formally ({\it Left}) and with our expanded definitions ({\it Right}). For the rest of the work we drop the ``-like" moniker and simply refer to the base name of the configuration.

Specifically, we do not expand the definition of Equilateral and Folded modes, as they are relatively smooth without the alteration. For every other mode, we relax the condition of equality in their formal definition to instead be a ratio lower limit. For example Isosceles modes formally impose $k_1=k_2$ but for our isosceles-like modes we impose $k_2/k_1 > 0.95$. We impose this same relaxation on all ``-like" modes. For the case of squeezed-like modes we introduce one further relaxation, the minimum ratio of $k_1/k_3 > 3$. These relaxations of the definitions not only smooth out the resulting functions but also allow for more angular information to be captured while still having a reasonable amount of overlap between different configurations.

\subsection{Spherical Bispectrum Bin Completeness: $k_{\rm sph,max}$}
An important consideration with the spherical bispectrum is that of bin completeness. Since the spherical bispectrum is a combination of various bispectrum elements with different ($k_1$, $k_2$, $k_3$) values, some elements may be put into a bin where other elements would in theory go but naively imposing a finite $k_{\rm max}$ could prevent those elements from being added into the bin. 

We can calculate the largest, in terms of scale, bin where this occurs by finding the smallest bispectrum element which is not captured with a cut at $k_{\rm max}$ and then compute which bin it would fall into. If we characterize $k_{\rm max}$ with its integer form $n_{k,\rm max} = k_{\rm max}/k_F$ ($k_F$ is the fundamental wavenumber) then depending on whether it is even or odd there are different forms for $n_{\rm sph, max}$, the integer form of the first incomplete bin, given by \reftab{evenodd}. There are a variety of different limits depending on the configuration.

To summarize, with the maximum wavenumber of $n_{k,{\rm max}}$, the spherical bispectrum is complete up to, but not including, $n_{\rm sph,max}$ indicated in \reftab{evenodd}.

\section{Result: The Range of Validity}\label{sec:scale}

\begin{table*}
\centering
\begin{tabular}{lcc}
\hline \hline
\multicolumn{1}{c|}{} &
\multicolumn{1}{|c|}{$n_{\rm sph,max}$ for even values of $n_{k, \rm max}$} &
\multicolumn{1}{|c}{$n_{\rm sph,max}$ for odd values of $n_{k, \rm max}$}
\\
\hline
\multicolumn{1}{l|}{General} & 
\multicolumn{1}{|c|}{$\sqrt{n_{k, \rm max}(n_{k, \rm max} + 2) + 4/3}/\sqrt{2}$} &
\multicolumn{1}{|c}{$(n_{k,\rm max}+1)/\sqrt{2}$}
\\
\multicolumn{1}{l|}{Folded} & 
\multicolumn{1}{|c|}{$(n_{k, \rm max} + 2)/\sqrt{2}$} &
\multicolumn{1}{|c}{$(n_{k, \rm max} + 1)/\sqrt{2}$}
\\
\multicolumn{1}{l|}{Elongated} & 
\multicolumn{1}{|c|}{$\sqrt{n_{k, \rm max}(n_{k, \rm max} + 2) + 4/3}/\sqrt{2}$} &
\multicolumn{1}{|c}{$(n_{k,\rm max}+1)/\sqrt{2}$}
\\
\multicolumn{1}{l|}{O-Isosceles} &
\multicolumn{1}{|c|}{$\sqrt{n_{k,\rm max}(n_{k,\rm max} + 8/3) + 2 }/\sqrt{2}$} &
\multicolumn{1}{|c}{$(n_{k,\rm max}+1)/\sqrt{2}$}
\\
\multicolumn{1}{l|}{Equilateral} & 
\multicolumn{1}{|c|}{$(n_{k,\rm max} + 1)$} &
\multicolumn{1}{|c}{$(n_{k,\rm max} + 1)$}
\\
\multicolumn{1}{l|}{Isosceles} & 
\multicolumn{1}{|c|}{$\sqrt{2 n_{k, \rm max} (n_{k,\rm max} + 2) + 3}/\sqrt{3}$} &
\multicolumn{1}{|c}{$\sqrt{2 n_{k, \rm max} (n_{k,\rm max} + 2) + 3}/\sqrt{3}$}
\\
\multicolumn{1}{l|}{Squeezed}  & 
\multicolumn{1}{|c|}{$\sqrt{2 n_{k, \rm max} (n_{k,\rm max} + 2) + 3}/\sqrt{3}$} &
\multicolumn{1}{|c}{$\sqrt{2 n_{k, \rm max} (n_{k,\rm max} + 2) + 3}/\sqrt{3}$}
\\
\hline\hline
\end{tabular}
\caption{The value of $n_{\rm sph,max} = k_{\rm sph,max}/k_f$ for each configuration when $n_{k, \rm max}$ is even/odd. This is calculated by finding the smallest uncaptured mode for each configuration and then calculating its $k_{\rm sph}$.}
\label{tab:evenodd}
\end{table*}

In this section, we present out results for the range of validity for both power and bispectrum at both LO and NLO. First, let us define the range of validity. 

While the general consensus in literature is that the nonlinear scale, or range of validity for the linear model, is where the NLO correction is a sizable fraction of the LO contribution, the exact definition of the nonlinear scale can vary throughout the literature. In this work, we use a more data-driven approach and compute a $\chi^2$-like statistic to determine $k_{\rm NL}$ where the simulation and theory models diverge beyond our accuracy threshold. Specifically, for a clustering statistic $F$, e.g. the power spectrum or bispectrum, we compute $n_{NL} = k_{NL}/k_f$ through
\be \label{eq:knl}
\frac{1}{n_{NL}}\sum_{i=1}^{n_{NL}} \frac{(F_{{\rm theory}, i} - F_{{\rm sim}, i})^2}{\sigma_{F_i,\rm sim}^2 + \sigma_{F_i, \rm theory}^2} < (1.5)^2 \, ,
\ee
where $\sigma_{F_i, \rm sim}$ is the statistical error on the mean of the clustering statistic measured from the simulations at point $i$. The $\sigma_{F_i,\rm theory}$ in denominator is an induced statistical error on the theory to take our accuracy threshold into account, defined as
\be
\sigma_{F_i,\rm theory} = \frac{{\rm acc}}{2} F_{\rm theory}(k_i) \, ,
\ee
where acc is our desired accuracy threshold, e.g. $1\%$ for the power spectrum. This theoretical error term, while only contributing acc/2 correction to cosmic variance, dominates over the shot-noise error near $k_{\rm NL}$. The contribution from the first five modes is suppressed by the usual cosmic variance and stays less than 1\% of the summation, which is dominated by contributions near $k_{\rm NL}$.
In effect, what this computes is the average deviation from our accuracy threshold in units of the error, i.e. in units of sigma, and finds where that crosses 1.5 sigma. 

A final note about our choice of 1.5 sigma is in order. Here, the 1.5 sigma constraint is somewhat arbitrary and was determined by empirical inspection, to match the $k_{\rm NL}$ we determine in \refeq{knl} and the values from the comparison plots such as \reffig{pkerrnl} and \reffig{bkerrnl}: using 1 sigma is too strict and 2 sigma is too lenient. The choice of threshold sigma, however, does not alter the results in any significant way and merely shifts the specific values of $k_{NL}$ up or down while maintaining the same general relationships. Typically the difference from 1 to 1.5 and 1.5 to 2 is about a $10\%$ difference at each step. For values of $k_{\rm NL}$ that do not fall exactly on a simulation bin center we use cubic spline interpolation to increase our precision.

\subsection{Power Spectrum}

\begin{figure*}
    \centering
    \includegraphics[width=.95\textwidth]{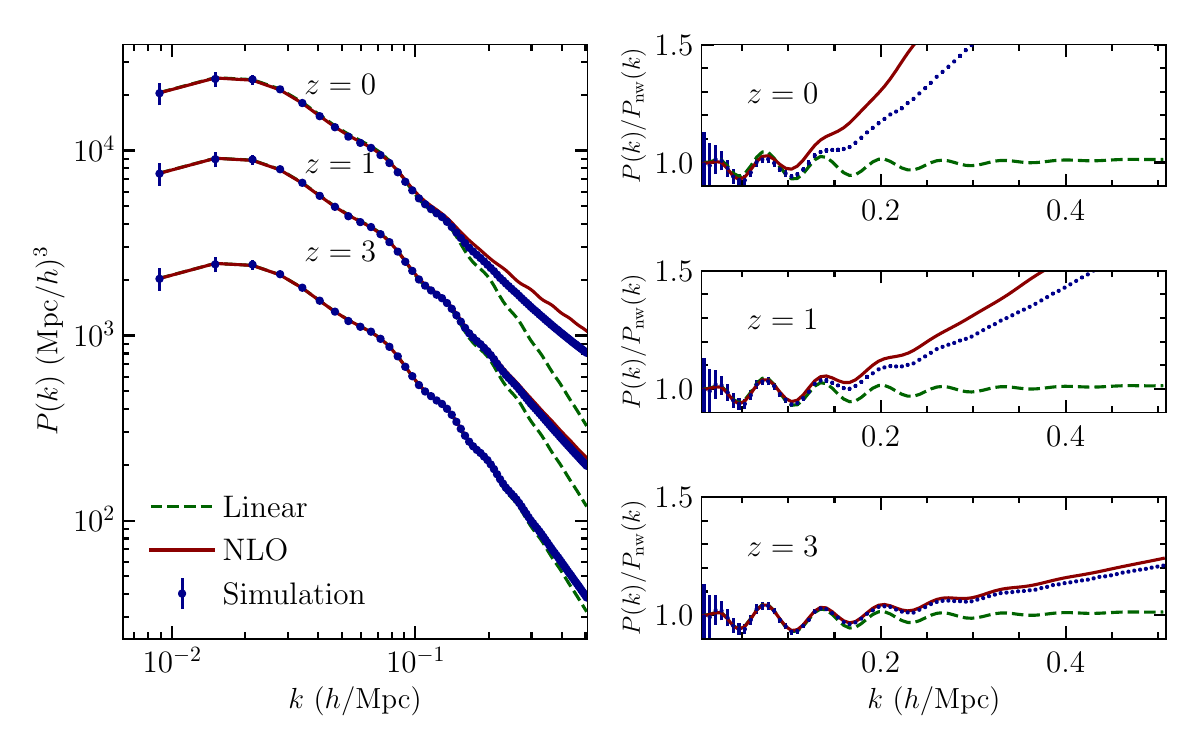}
    \caption{({\it Left}): The power spectrum for all three of our models at all three plotting redshifts (0, 1, 3). We plot all three redshifts on the same plot since the scaling makes them mostly distinct. The largest valued set of power spectra is the $z=0$ set, while the smallest is the $z=3$ set. It can clearly be seen by eye how much better the high redshift curves match at smaller scales compared to the $z=0$ set. \\
    ({\it Right}): The various power spectrum models divided by the no-wiggle power spectrum from \cite{Eisenstein_1998}. The redshifts follow the same pattern as on the left. Note that for both plots the errors are actually five-sigma errors to enhance visibility.}
    \label{fig:bigpk}
\end{figure*}

To visualize the differences between the two (LO and LO+NLO) models and the N-body result we present a plot of all three curves at all three plotting redshifts ($z=0$, $1$, $3$) in \reffig{bigpk}. To facilitate visualization of the nonlinearities in BAO, we also divide the curves by the no-wiggle power spectrum from \cite{Eisenstein_1998} on the right panel of \reffig{bigpk}.

 Here, we see excellent agreement between everything on large scales, with the small-scale agreement being highly dependent on redshift, as expected \citep{Jeong/Komatsu:2006}. The non-wiggle plots emphasize the baryon acoustic oscillation (BAO) feature and show that LO+NLO perturbation theory prediction can accurately capture the nonlinearities in BAO at $z=3$, while failing at lower redshifts ($z=0$, $1$).

\begin{figure*}
    \centering
    \includegraphics[width=.49\textwidth]{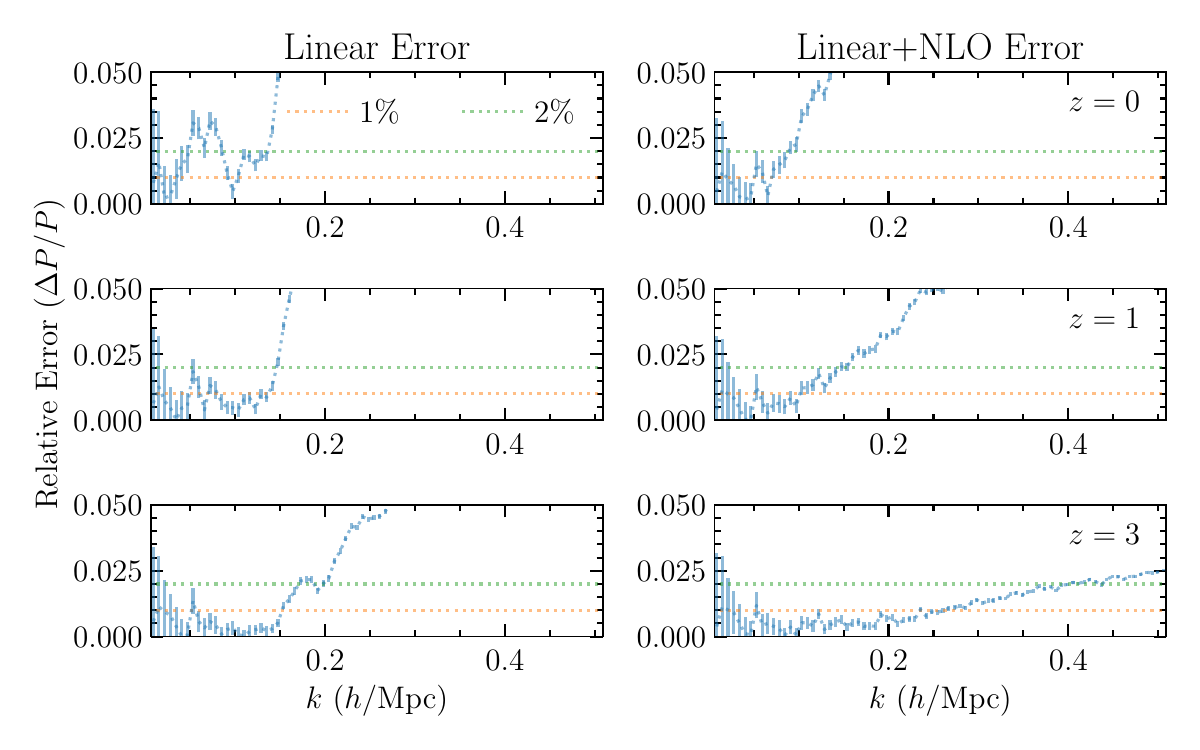}
    \includegraphics[width=.49\textwidth]{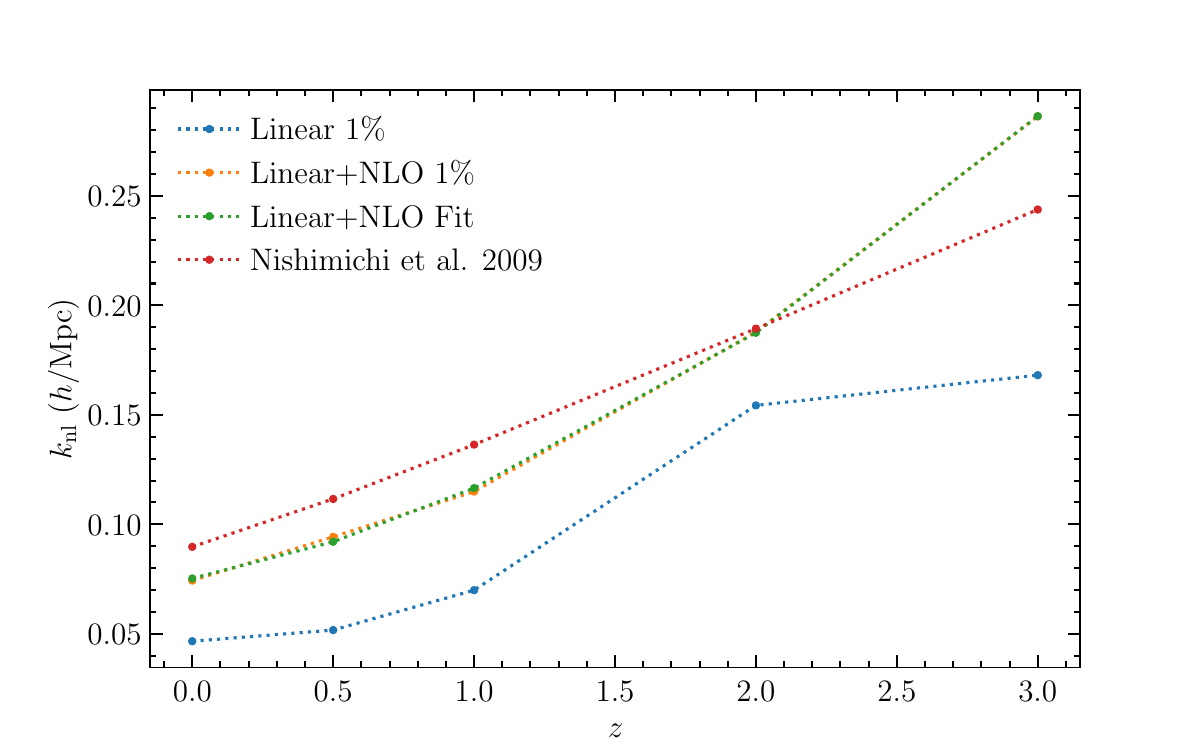}
    \caption{({\it Left}): The relative error for each model compared to the simulation at three different redshifts, from top to bottom $z=0$, $z=1$, $z=3$. Left: The error for the three linear models. Here we see significant redshift evolution, with the first point inconsistent with 1\% going from $\sim$ 0.05 $h$/Mpc at $z=0$ to $\sim$ 0.15 $h$/Mpc at $z=3$. Right: The error for the three NLO models sees a similar redshift evolution, going from diverging at $\sim$ 0.09 $h$/Mpc at $z=0$ to $\sim$ 0.3 $h$/Mpc at $z=3$. It also remains consistently better than just the linear model. Overall we see significant accuracy improvement from both increasing the linear model to NLO and from increasing redshift. \\
 ({\it Right}):    The value of $k_{\rm NL}$ for each redshift using both linear and L+NLO models. These values were computed using \refeq{knl}. We see a clear increase with redshift for both linear and NLO models, expected for perturbative methods. We fit a simple model to our results, given by \refeq{pknlfit} that we plot alongside the model given by \cite{Nishimichi_2009}. Note that although the form of the model is cosmology independent it was calibrated using a significantly different cosmology so differences of these levels are not unexpected. The same caution should be taken when applying our fit model.}
    \label{fig:pkerrnl}
\end{figure*}

We also show the error plots for our three plotting redshifts ($z=0$, $1$, $3$) in the left panel of \reffig{pkerrnl} along with overplotted lines to highlight 1\% and 2\% residuals. We see the expected redshift evolution that both LO and NLO perturbation theory predictions become more accurate at higher redshifts. Corresponding $k_{\rm NL}$ values as we compute following \refeq{knl} can be seen for all redshifts and models in the right panel of \reffig{pkerrnl}. Note that there is a sharp discontinuity in the ``Linear 1\%'' model. This sudden increase of $k_{\rm NL}$ is due to the first BAO wiggle being inconsistent to 1\% at low redshift but consistent at high redshift. Besides that, the scaling shows the expected behavior with slow increases on either side of the linear discontinuity and steep increases at each redshift step for the LO+NLO model.

We fit a simple model to the results of our calculation for the LO+NLO results and get an expression
\be \label{eq:pknlfit}
\frac{k_{\rm NL}(z)}{h/{\rm Mpc}} = 0.0502 + 0.0251 \left(\frac{D(z)}{D(0)} \right)^{-1.939}
\ee
where $D(z)$ is the linear growth factor. We do not fit a similar curve for the linear model due to the  discontinuity caused by the first BAO wiggle. Along with our basic fit model we also look at the model described in \cite{Nishimichi_2009}. This takes the form
\be
\frac{k_{\rm NL}(z)^2}{6\pi^2}\int_0^{k_{NL}(z)} P_L(q,z) \, {\rm d}q < 0.18
\ee
Where the constant on the right hand side varies depending on the model, here we use their $C_{1\%}^{\rm SPT}$. We find these $k_{NL}$ values for our L+NLO model and plot it alongside our data and model in the right panel of \reffig{pkerrnl}. Our fitted model is quite consistent with our results, particularly at higher redshifts. The \cite{Nishimichi_2009} model has similar behaviour to our model, albeit with different scaling at high redshift, not unexpected for a model that was calibrated with a significantly different cosmology, WMAP3 \citep{WMAP32007} ($\Omega_m=0.234$ $\sigma_8=0.76$) vs Planck18 \citep{Planck2018} ($\Omega_m=0.315$ $\sigma_8=0.811$). The difference suggests that care should be taken before applying \refeq{pknlfit} that, even though it is written in a form to be generalizable to other cosmologies, is likely to have errors when applied to moderately different cosmologies.

\subsection{Bispectrum}

\begin{figure*}
    \centering
    \includegraphics[width=0.8\textwidth]{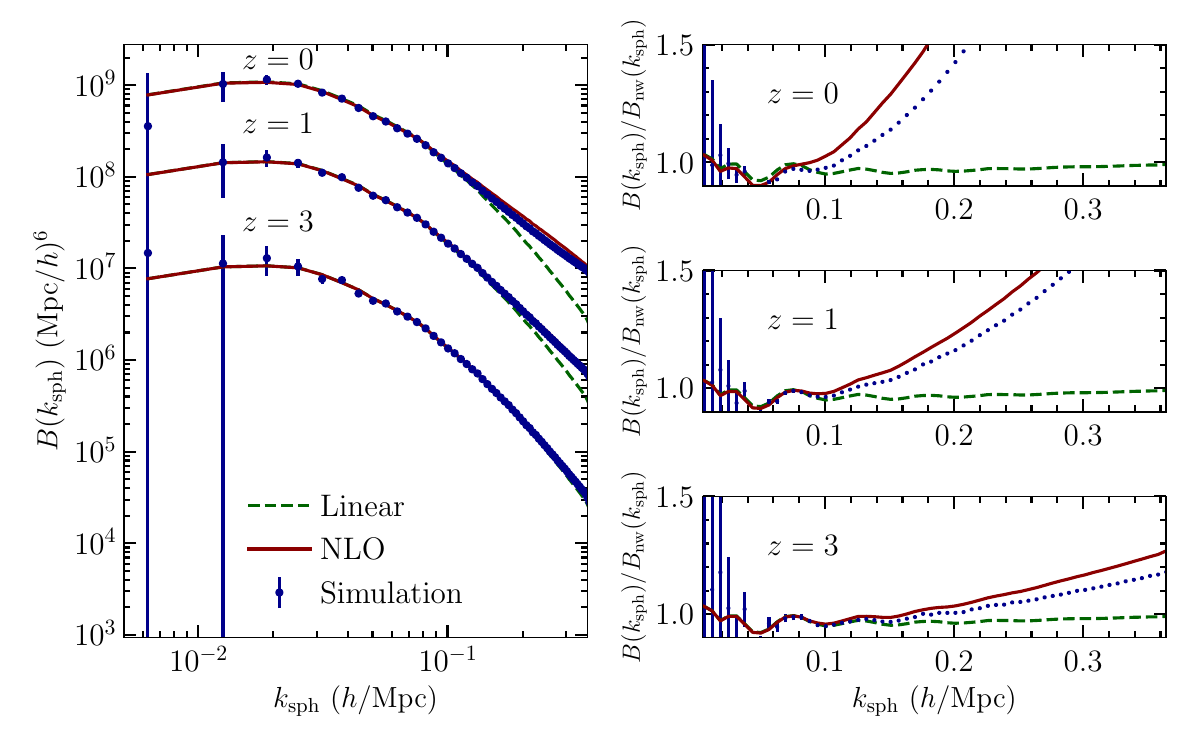}
    \caption{Left: The bispectrum for all three of our models at all three plotting redshifts (0, 1, 3). We plot all three redshifts on the same plot since the scaling makes them mostly distinct. The largest valued set of bispectra is the z=0 set, while the smallest is the z=3 set. It can clearly be seen by eye how much better the high redshift curves match at smaller scales compared to the z=0 set. Right: The various bispectrum models divided by the no-wiggle bispectrum, the LO bispectrum generated using the no-wiggle power spectrum from \cite{Eisenstein_1998}. The redshifts follow the same pattern as on the left.}
    \label{fig:bigbk}
\end{figure*}

Similarly to the power-spectrum case, here we present spherical bispectrum plots for all redshifts ($z=0$, $1$, $3$) and models (LO and LO+NLO) in \reffig{bigbk}. We see very similar behavior to the power spectrum, one of our motivations for introducing the spherical-bispectrum visualization technique, with the LO+NLO model having better overall agreement and the agreement becomes better with redshift. Since the full bispectrum does not have a single scale associated with each element, we define the range of validity $k_{\rm sph, NL}$ based off the spherical bispectrum. The use of $k_{\rm sph, NL}$ allows us to clearly analyze a single effective scale for deviation of the two models from N-body data.

\begin{figure*}
    \centering
    \includegraphics[width=.75\textwidth]{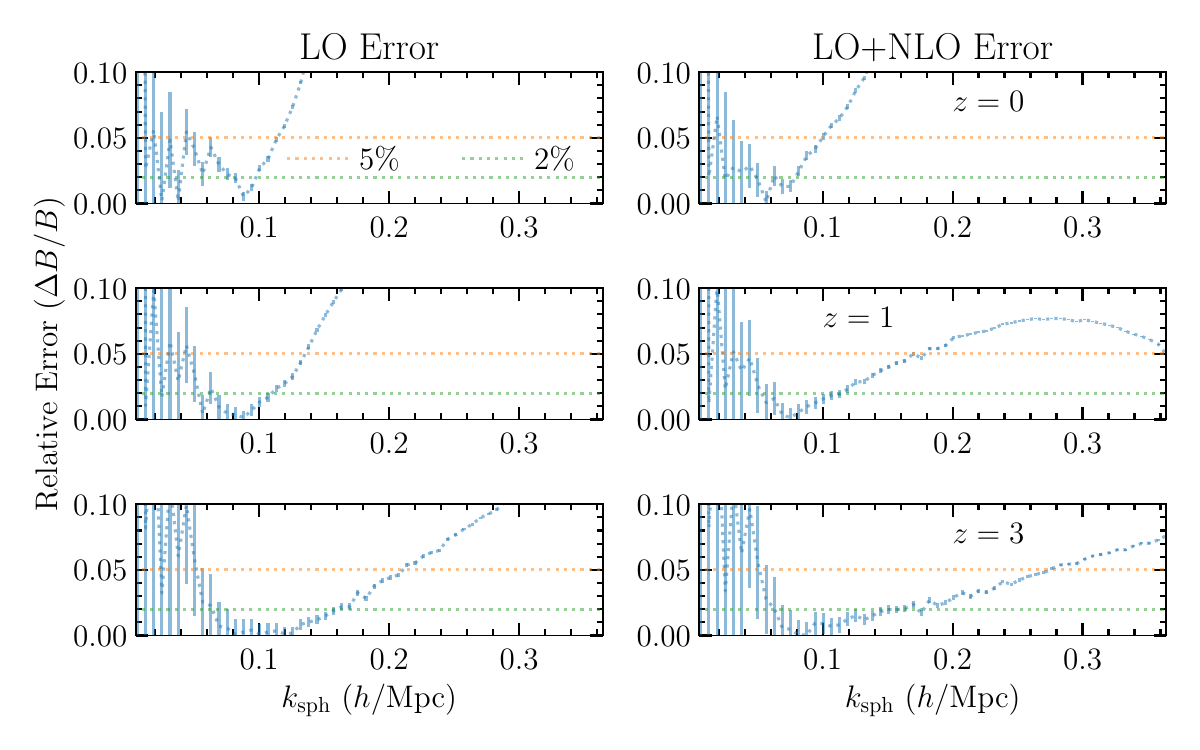}
    \includegraphics[width=.99\textwidth]{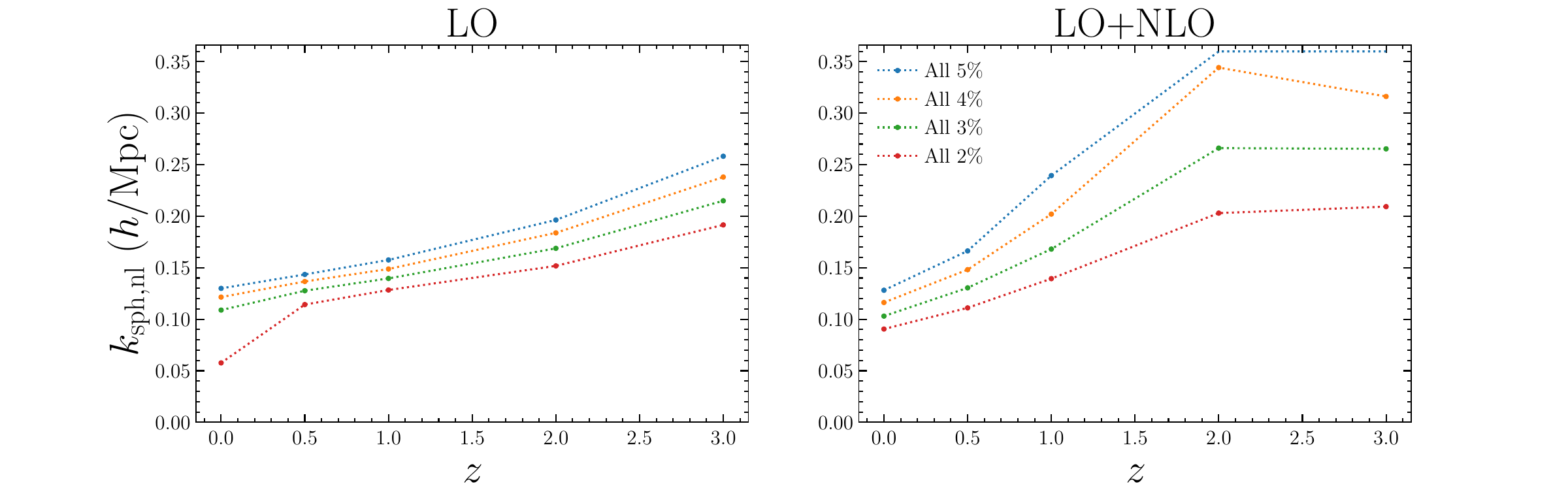}
    \caption{({\it Top}): The relative error for each model compared to the simulation at three different redshifts, from top to bottom $z=0$, $z=1$, $z=3$. Left: The error for the three LO models. Here we see significant redshift evolution, with the first point inconsistent with 5\% going from $\sim$ 0.2 $h$/Mpc at $z=0$ to $\sim$ 0.375 $h$/Mpc at $z=3$. Right: The error for the three LO+NLO models sees a similar redshift evolution, going from diverging at $\sim$ 0.2 $h$/Mpc at $z=0$ to $\sim$ 0.5 $h$/Mpc at $z=3$. It also remains consistently better than just the linear model. Overall we see significant accuracy improvement from both increasing the LO model to LO+NLO and from increasing redshift.\\
 ({\it Bottom}): The value of $k_{\rm sph, nl}$ found using \refeq{knl} with the spherical bispectrum at four different threshold values. Left: The LO bispectrum results at four different accuracy thresholds. In general we see a smooth increase in $k_{\rm sph, nl}$ with both redshift and accuracy threshold. Right: The results for the LO+NLO model at four accuracy thresholds. Generally the results are similar to the LO results, except for a slight flattening/decrease between redshifts 2 and 3. The 5\% results also hit the maximum value for which $k_{\rm sph}$ is complete, implying it is a lower bound.}
    \label{fig:bkerrnl}
\end{figure*}

We essentially use \refeq{knl} but with $B(k_{\rm sph})$ as $F$, and with a few different cutoff errors, 2-5\%, instead of 1\% as for the power spectrum since both observational and statistical errors are larger for the bispectrum. This is because the bispectrum is a higher-order statistic, involving a larger number of individual measurements. We present relative error plots of both LO and LO+NLO perturbative bispectrum models at our three plotting redshifts ($z=0$, $1$, $3$) in the top panels of \reffig{bkerrnl} with overplotted lines to highlight 2\% and 5\% errors, the bounds of our analysis. We again see a similar pattern with LO+NLO having better accuracy than LO, and the accuracy also increasing with redshift for both models. Our final results for the total bispectrum $k_{\rm sph, nl}$ are presented in the \reffig{bkerrnl}. 
Due to the difference between $k_{\rm sph}$ and $k$ and the nature of $B_{\rm sph}$ itself, it is hard to make a direct comparison between the bottom panel of \reffig{bkerrnl} and the right panel of \reffig{pkerrnl}. Nevertheless, we can read off the trend that $k_{\rm NL}$ with 1\% accuracy in power spectrum corresponds to $k_{\rm sph, NL}$ with 3\% accuracy in bispectrum.

We can also perform this comparison between the theory calculation and N-body results for specific configurations to disentangle the averaging effect in $B_{\rm sph}$. This study can also clarify that the accuracy of PT's bispectrum modeling depends on the configuration. To compute $k_{\rm sph, NL}^{\rm (config)}$ at each cutoff percentage, we again use \refeq{knl} but with the bispectrum filtered to a specific configuration, instead of the bispectrum of all modes. We repeat the calculation for each model and with all six configurations. 

\begin{figure*}
    \centering
    \includegraphics[width=0.99\textwidth]{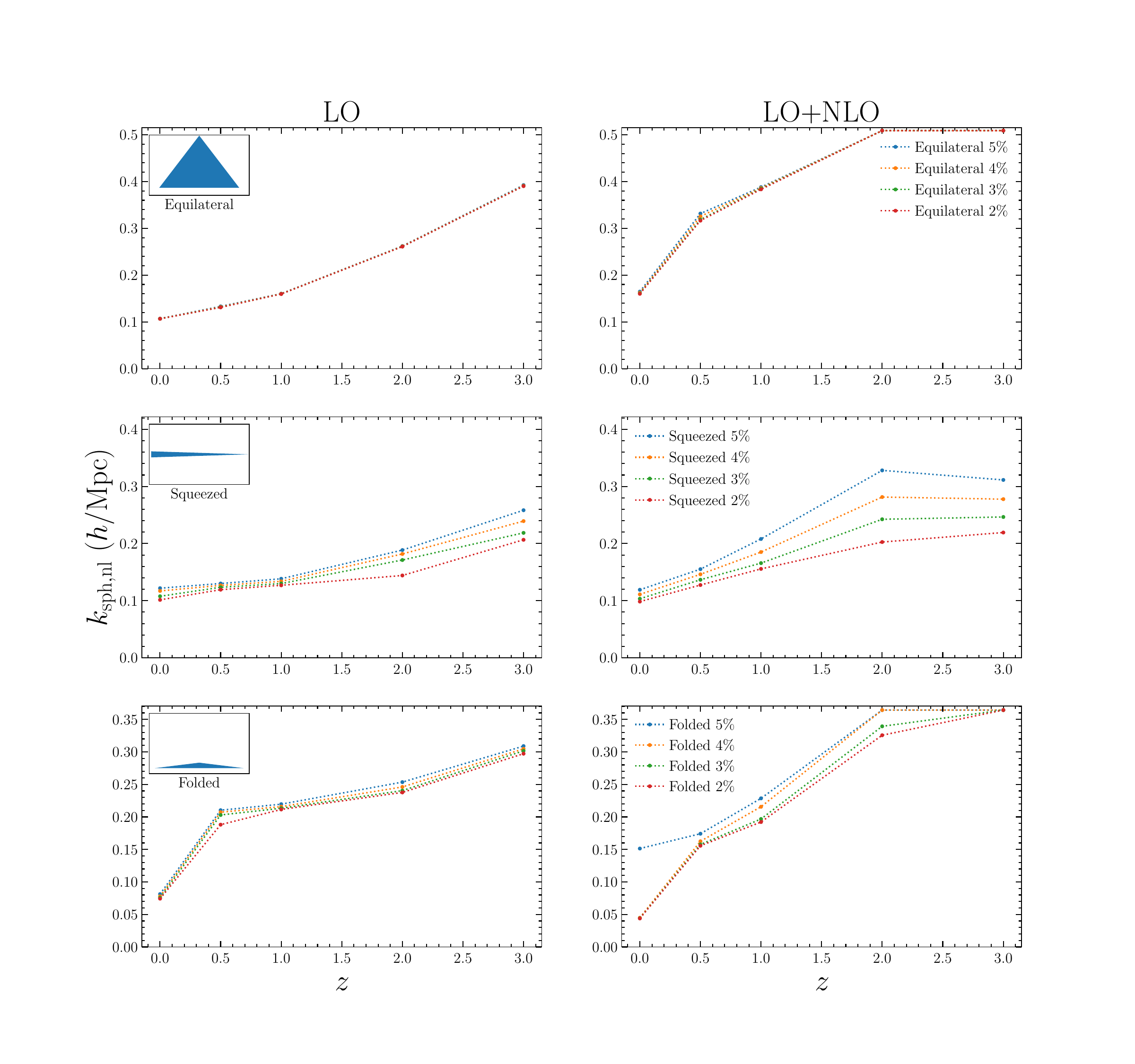}
    \caption{The value of $k_{\rm sph, nl}$ found using \refeq{knl} with the spherical bispectrum, filtered to a specific configuration, at four different threshold values. Left: The results using the LO bispectrum model. Inset in each plot is a representative triangle showing the definition of the configuration. The equilateral results show that the threshold accuracy, within our bounds, does not change $k_{\rm sph, nl}$, signifying that the loss of accuracy is abrupt, but has a reasonable redshift scaling. Squeezed configurations seem to behave as expected, although the dependence on accuracy threshold is quite weak. Folded configurations seems to have little accuracy dependence, and experiences a significant jump from redshift 0 to 0.5. Right: The results for the LO+NLO model. At NLO Equilateral configurations retain their lack of dependence on accuracy threshold. It also hits the maximum of the data at both redshift 2 and 3, representing lower bounds. NLO Squeezed configurations have a much stronger accuracy dependence, with an added flattening between redshift 2 and 3. At NLO Folded configurations have more of a dependence on accuracy, hitting the maximum value at redshift 2, for some thresholds, and redshift 3, for all thresholds.}
    \label{fig:bknlconfig1}
\end{figure*}

\begin{figure*}
    \centering
    \includegraphics[width=0.99\textwidth]{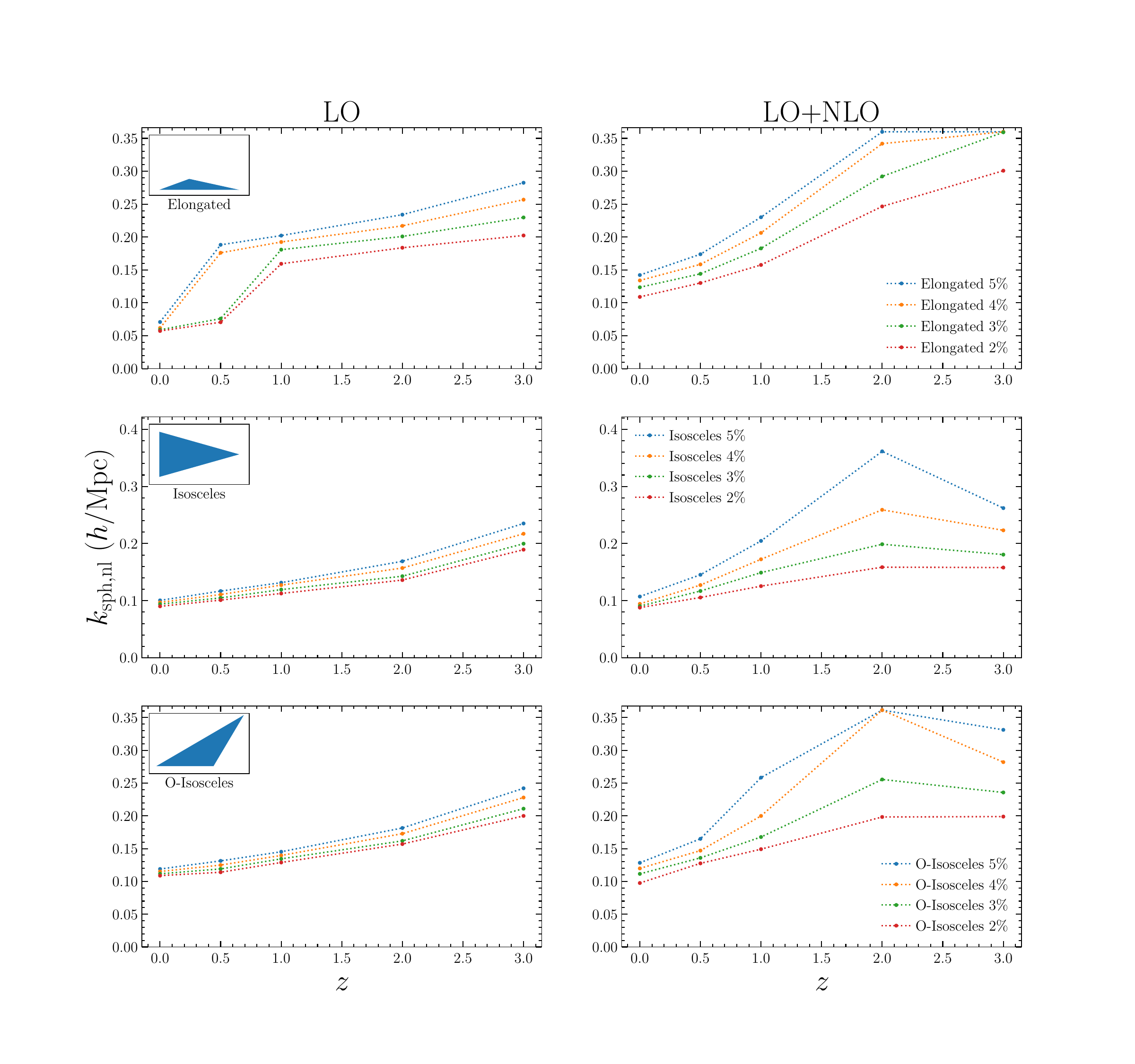}
    \caption{The value of $k_{\rm sph, nl}$ found using \refeq{knl} with the spherical bispectrum, filtered to a specific configuration, at four different threshold values. Left: The results using the LO bispectrum model. Inset in each plot is a representative triangle showing the definition of the configuration. The elongated results show expected redshift and accuracy scaling, except between redshift 0 and 1 where the higher thresholds scale rapidly while the lower thresholds more slowly. The Isosceles configuration results show reasonable scaling with redshift, with a weak dependence on accuracy threshold. Similarly the Obtuse-Isosceles results are very similar to the Isosceles results. Right: The results for the LO+NLO model. At NLO Elongated configurations display the expected scaling, with some of the redshift 2 and 3 results hitting the maximum value, implying a lower bound. The NLO Isosceles configurations have a strange feature where there is a significant drop from redshift 2 to 3, at least for 5\% accuracy. There are many potential causes for this, an unlikely statistical fluctuation, an unexpected binning error, or many other possibilities. At NLO Obtuse-Isosceles behaves similarly to Isosceles, with a drop from redshift 2 to 3, although it does hit the maximum value at redshift 2. }
    \label{fig:bknlconfig2}
\end{figure*}

We show the results for the each configuration in \reffig{bknlconfig1} and \reffig{bknlconfig2}. Here we see that most configurations behave in a similar way, LO+NLO is more accurate than just LO, often by a significant amount, and accuracy improves dramatically at higher redshifts with our accuracy threshold having the expected minor scaling. There are a number of mildly anomalous results in that there are a number of points where LO is equally as accurate as LO+NLO, in the All, Isosceles, and Obtuse-Isosceles plots we see a slight decrease in accuracy going from $z=2$ to $z=3$. We attribute these features to either be statistical fluctuations or a consequence of imperfect binning corrections, some configurations are significantly more dependent on the binning effect than others \citep{Bernardeau_2012,eggemeier2021testing}. Due to the inaccuracy of our large-scale bispectrum measurements we leave the investigation of this problem for future work. A few other features of interest in the plot are the Equilateral curves having little variation with threshold. This represents the Equilateral models being very accurate with a sudden divergence from the model causing all the thresholds to be passed at the same point. There are also a few discontinuous points with the same cause as for the LO power spectrum (the blue line in the right panel of \reffig{pkerrnl}), a feature that isn't captured at lower redshifts accurately that falls under the threshold as the redshift increases.

\section{Conclusion}\label{sec:conclusion}

The galaxy power spectrum and bispectrum are some of the main summary statistics for studying large-scale structure (LSS) cosmology. Unlike the CMB anisotropies, where $\Delta T/T\simeq 10^{-5}$ is small, linear perturbation theory fails to describe late-time cosmic density field whose r.m.s. fluctuation is of order unity on relevant scales. For LSS cosmology, therefore, it is essential to model the nonlinear evolution of the power spectrum and bispectrum, and standard perturbation theory (SPT) provides an analytical expression for the nonlinear contributions, we call NLO (next-to-leading order) in this paper. To properly exploit the power of an analytic theory a firm grasp on the systematics and region of validity are required. In this paper, we analyze the range of validity of SPT by comparing the analytical calculation to a suite of N-body simulations.

We first compare the SPT power spectrum at both leading order (LO) and NLO to the results of one hundred high-resolution run of Quijote simulation and derive a range of validity, or $k_{\rm NL}$ the maximum wavenumber for which the SPT theory is accurate to 1\% in comparison to the N-body result. For the LO+NLO SPT matter power spectrum the we find a fitting formula for $k_{\rm NL}[D(z)]$ in \refeq{pknlfit}.

To carry out a similar analysis for the matter bispectrum, we have introduced a new visualization technique, the spherical bispectrum. The spherical bispectrum allows for plotting and analysis of the bispectra as a one-dimensional function of of $k_{\rm sph}$ that show the length-scale dependence. At the same time, we can show the triangular-configuration dependence by fixing the range of $\theta_{\rm sph}$ and $\phi_{\rm sph}$ to define the configuration. This scheme, therefore, retains {\it all} information in the bispectrum. 
By using this technique we find a good proxy for the maximum wavenumber for the bispectrum in \reffig{bkerrnl}.

These results allow for more accurate assessment for the analytic theory's constraining power by properly limiting the smallest scale above which each theory can accurately model the nonlinearities. That is, combined with an accurate covariance matrix \citep{Villaescusa_Navarro_2020}, since on small scales there is often significant mixing between different wave modes which reduces the information gain by including more modes, one can achieve very accurate predicted constraints. We leave that as a future work.

One issue that the bispectrum analysis must resolve, but we have only discussed briefly in this paper, is a more detailed analysis for proper binning of the continuous theory models, to confront the theory calculation with the data. There are many recent techniques attempting to alleviate the computational cost while maintaining good accuracy \citep{Bernardeau_2012,Sefusatti_2010,Oddo_2020,eggemeier2021testing}, but to our knowledge no comprehensive study has been done to compare the various methods while dealing with other potential issues such as interpolation scheme and bin size. Unlike for the power-spectrum analysis where binning the theory prediction causes only a minor problem for the first few modes, this is a key issue for the bispectrum that still requires more systematic study for accurate data analysis.

Of course, the range of validity analysis must also be done for the next-order clustering statistic, the trispectrum or four point function. The challenge with this type of extension is the significant increase in estimation cost, estimating a trispectrum with similar bin size and max wavenumber as a bispectrum takes $\sim$10 times as long \citep{Tomlinson_2019}. Similarly, because the cost of binning a continuous theory is proportional to the cost of estimation the binning problem is even larger for analytic trispectrum theories. All of these issues compound with the high computational cost of evaluating the NLO trispectrum, likely limiting study to only the cheap to compute LO part of the trispectrum. On the other hand, the trispectrum is expected to be an important probe of primordial non-Gaussanities \citep{Gualdi_2021,Gualdi_2021b}, as well as parity-violating new physics \citep{Cahn2021}. The LSS analysis with trispectrum has just begun recently \citep{Philcox2021,Gualdi2022}, and the high cost of including the trispectrum in future analysis demands a detailed study of the information contained in the trispectrum, similar to what \cite{Hahn_2020} did for the bispectrum, is needed to assess how practical the trispectrum is to use.

In this work, we have focused solely on the analytical SPT models that do not include any {\it free} parameters. Most extended models are built off a foundation of SPT with a couple of free parameters. Therefore, more advanced models such as Effective Field Theory (EFT) \citep{Baumann2012,Carrasco_2012,Carrasco_2014,Carrasco_2014b,Hertzberg2014,Senatore_2015}, Regularized Perturbation Theory (RegPT) \citep{Taruya2012}, Renormalized Lagrangian Perturbation Theory (RLPT) \citep{Matsubara2008,Rampf_2012}, Galilean-invariant Renormalized Perturbation Theory (gRPT) \citep{Eggemeier/etal:2020} or the general bias expansion \citep{Desjacques_2018review,Desjacques_2018,Tomlinson_2020} can use this type of study as the basis for either their own model limitations or to seed a more detailed study of each model. Generically, the bounds $k_{\rm NL}$ in this work should serve as lower bounds for these more advanced methods which often have different corrections to improve accuracy. For example, EFT often uses a counterterm like $\Delta P \propto k^2P_L$ which based off the results in this work, \reffig{pkerrnl}, should be a good choice to improve accuracy. Note that the same argument applies to the stochastic bias parameter in the general bias method. For a comparison of the range of validity of a few different methods at $z=1$, see \cite{Alkhanishvili_2022}.

\section*{Acknowledgements}
The authors would like to thank Francisco Villaescusa-Navarro for assistance with data and computing access.
We would also like to thank the whole Quijote team for making the simulation suite available and well documented.
The author acknowledge that the work reported on in this paper was substantially performed using the Princeton Research Computing resources at Princeton University which is consortium of groups led by the Princeton Institute for Computational Science and Engineering (PICSciE) and Office of Information Technology's Research Computing.
This work was supported at Pennsylvania State University by NASA ATP program (80NSSC18K1103) and NASA FINESST (80NSSC22K1751).

\bibliographystyle{JHEP}
\bibliography{main.bib}

\providecommand{\href}[2]{#2}\begingroup\raggedright\begin{thebibliography}{10}

\bibitem{dore2015cosmology}
O.~Doré, J.~Bock, M.~Ashby, P.~Capak, A.~Cooray, R.~de~Putter et~al.,
  \emph{Cosmology with the spherex all-sky spectral survey},  2015.

\bibitem{hill2008hobbyeberly}
G.J.~Hill, K.~Gebhardt, E.~Komatsu, N.~Drory, P.J.~MacQueen, J.~Adams et~al.,
  \emph{The hobby-eberly telescope dark energy experiment (hetdex): Description
  and early pilot survey results},  2008.

\bibitem{laureijs2011euclid}
R.~{Laureijs}, J.~{Amiaux}, S.~{Arduini}, J.L.~{Augu{\`e}res}, J.~{Brinchmann},
  R.~{Cole} et~al., \emph{{Euclid Definition Study Report}}, {\emph{arXiv
  e-prints} (2011) arXiv:1110.3193}
  [\href{https://arxiv.org/abs/1110.3193}{{\ttfamily 1110.3193}}].

\bibitem{levi2013desi}
M.~Levi, C.~Bebek, T.~Beers, R.~Blum, R.~Cahn, D.~Eisenstein et~al., \emph{The
  desi experiment, a whitepaper for snowmass 2013},  2013.

\bibitem{lsstdarkenergysciencecollaboration2012large}
L.D.E.S.~Collaboration, \emph{Large synoptic survey telescope: Dark energy
  science collaboration},  2012.

\bibitem{maartens2015cosmology}
R.~Maartens, F.B.~Abdalla, M.~Jarvis and M.G.~Santos, \emph{Cosmology with the
  ska -- overview},  2015.

\bibitem{spergel2015widefield}
D.~Spergel, N.~Gehrels, C.~Baltay, D.~Bennett, J.~Breckinridge, M.~Donahue
  et~al., \emph{Wide-field infrarred survey telescope-astrophysics focused
  telescope assets wfirst-afta 2015 report},  2015.

\bibitem{Takada_2014}
M.~Takada, R.S.~Ellis, M.~Chiba, J.E.~Greene, H.~Aihara, N.~Arimoto et~al.,
  \emph{Extragalactic science, cosmology, and galactic archaeology with the
  subaru prime focus spectrograph},
  \href{https://doi.org/10.1093/pasj/pst019}{\emph{Publications of the
  Astronomical Society of Japan} {\bfseries 66} (2014) R1}.

\bibitem{Schmittfull_2016}
M.~Schmittfull, Z.~Vlah and P.~McDonald, \emph{Fast large scale structure
  perturbation theory using one-dimensional fast fourier transforms},
  \href{https://doi.org/10.1103/physrevd.93.103528}{\emph{Physical Review D}
  {\bfseries 93} (2016) }.

\bibitem{McEwen_2016}
J.E.~McEwen, X.~Fang, C.M.~Hirata and J.A.~Blazek, \emph{Fast-pt: a novel
  algorithm to calculate convolution integrals in cosmological perturbation
  theory}, \href{https://doi.org/10.1088/1475-7516/2016/09/015}{\emph{Journal
  of Cosmology and Astroparticle Physics} {\bfseries 2016} (2016) 015–015}.

\bibitem{slepian2018decoupling}
Z.~Slepian, \emph{On decoupling the integrals of cosmological perturbation
  theory},  2018.

\bibitem{Schmittfull_2016two}
M.~Schmittfull and Z.~Vlah, \emph{Reducing the two-loop large-scale structure
  power spectrum to low-dimensional, radial integrals},
  \href{https://doi.org/10.1103/physrevd.94.103530}{\emph{Physical Review D}
  {\bfseries 94} (2016) }.

\bibitem{Simonovi__2018}
M.~Simonović, T.~Baldauf, M.~Zaldarriaga, J.J.~Carrasco and J.A.~Kollmeier,
  \emph{Cosmological perturbation theory using the fftlog: formalism and
  connection to qft loop integrals},
  \href{https://doi.org/10.1088/1475-7516/2018/04/030}{\emph{Journal of
  Cosmology and Astroparticle Physics} {\bfseries 2018} (2018) 030–030}.

\bibitem{Scoccimarro_2004}
R.~Scoccimarro, \emph{Redshift-space distortions, pairwise velocities, and
  nonlinearities},
  \href{https://doi.org/10.1103/physrevd.70.083007}{\emph{Physical Review D}
  {\bfseries 70} (2004) }.

\bibitem{Taruya_2010}
A.~Taruya, T.~Nishimichi and S.~Saito, \emph{Baryon acoustic oscillations in
  2d: Modeling redshift-space power spectrum from perturbation theory},
  \href{https://doi.org/10.1103/physrevd.82.063522}{\emph{Physical Review D}
  {\bfseries 82} (2010) }.

\bibitem{Zheng_2010}
Z.~Zheng, R.~Cen, H.~Trac and J.~Miralda-Escudé, \emph{Radiative transfer
  modeling of ly$\alpha$ emitters. ii. new effects on galaxy clustering},
  \href{https://doi.org/10.1088/0004-637x/726/1/38}{\emph{The Astrophysical
  Journal} {\bfseries 726} (2010) 38}.

\bibitem{Hirata_2009}
C.M.~Hirata, \emph{Tidal alignments as a contaminant of redshift space
  distortions},
  \href{https://doi.org/10.1111/j.1365-2966.2009.15353.x}{\emph{Monthly Notices
  of the Royal Astronomical Society} {\bfseries 399} (2009) 1074–1087}.

\bibitem{Blazek_2015}
J.~Blazek, Z.~Vlah and U.~Seljak, \emph{Tidal alignment of galaxies},
  \href{https://doi.org/10.1088/1475-7516/2015/08/015}{\emph{Journal of
  Cosmology and Astroparticle Physics} {\bfseries 2015} (2015) 015–015}.

\bibitem{Eggemeier_2019}
A.~Eggemeier, R.~Scoccimarro and R.E.~Smith, \emph{Bias loop corrections to the
  galaxy bispectrum},
  \href{https://doi.org/10.1103/physrevd.99.123514}{\emph{Physical Review D}
  {\bfseries 99} (2019) }.

\bibitem{Desjacques_2018review}
V.~Desjacques, D.~Jeong and F.~Schmidt, \emph{Large-scale galaxy bias},
  \href{https://doi.org/10.1016/j.physrep.2017.12.002}{\emph{Physics Reports}
  {\bfseries 733} (2018) 1–193}.

\bibitem{Desjacques_2018}
V.~Desjacques, D.~Jeong and F.~Schmidt, \emph{The galaxy power spectrum and
  bispectrum in redshift space},
  \href{https://doi.org/10.1088/1475-7516/2018/12/035}{\emph{Journal of
  Cosmology and Astroparticle Physics} {\bfseries 2018} (2018) 035–035}.

\bibitem{Tomlinson_2020}
J.~Tomlinson, H.S.G.~Gebhardt and D.~Jeong, \emph{Fast calculation of the
  nonlinear redshift-space galaxy power spectrum including selection bias},
  \href{https://doi.org/10.1103/physrevd.101.103528}{\emph{Physical Review D}
  {\bfseries 101} (2020) }.

\bibitem{Ivanov_2020}
M.M.~Ivanov, M.~Simonović and M.~Zaldarriaga, \emph{Cosmological parameters
  from the boss galaxy power spectrum},
  \href{https://doi.org/10.1088/1475-7516/2020/05/042}{\emph{Journal of
  Cosmology and Astroparticle Physics} {\bfseries 2020} (2020) 042–042}.

\bibitem{Smith_2008}
R.E.~Smith, R.K.~Sheth and R.~Scoccimarro, \emph{Analytic model for the
  bispectrum of galaxies in redshift space},
  \href{https://doi.org/10.1103/physrevd.78.023523}{\emph{Physical Review D}
  {\bfseries 78} (2008) }.

\bibitem{Rampf_2012}
C.~Rampf and Y.Y.~Wong, \emph{Lagrangian perturbations and the matter
  bispectrum ii: the resummed one-loop correction to the matter bispectrum},
  \href{https://doi.org/10.1088/1475-7516/2012/06/018}{\emph{Journal of
  Cosmology and Astroparticle Physics} {\bfseries 2012} (2012) 018–018}.

\bibitem{Assassi_2014}
V.~Assassi, D.~Baumann, D.~Green and M.~Zaldarriaga, \emph{Renormalized halo
  bias}, \href{https://doi.org/10.1088/1475-7516/2014/08/056}{\emph{Journal of
  Cosmology and Astroparticle Physics} {\bfseries 2014} (2014) 056–056}.

\bibitem{Bernardeau_2012}
F.~Bernardeau, M.~Crocce and R.~Scoccimarro, \emph{Constructing regularized
  cosmic propagators},
  \href{https://doi.org/10.1103/physrevd.85.123519}{\emph{Physical Review D}
  {\bfseries 85} (2012) }.

\bibitem{Lazanu_2016}
A.~Lazanu, T.~Giannantonio, M.~Schmittfull and E.~Shellard, \emph{Matter
  bispectrum of large-scale structure: Three-dimensional comparison between
  theoretical models and numerical simulations},
  \href{https://doi.org/10.1103/physrevd.93.083517}{\emph{Physical Review D}
  {\bfseries 93} (2016) }.

\bibitem{Angulo_2015}
R.E.~Angulo, S.~Foreman, M.~Schmittfull and L.~Senatore, \emph{The one-loop
  matter bispectrum in the effective field theory of large scale structures},
  \href{https://doi.org/10.1088/1475-7516/2015/10/039}{\emph{Journal of
  Cosmology and Astroparticle Physics} {\bfseries 2015} (2015) 039–039}.

\bibitem{eggemeier2021testing}
A.~Eggemeier, R.~Scoccimarro, R.E.~Smith, M.~Crocce, A.~Pezzotta and
  A.G.~Sánchez, \emph{Testing one-loop galaxy bias: joint analysis of power
  spectrum and bispectrum},  2021.

\bibitem{McCullagh/Jeong/Szalay:2015}
N.~McCullagh, D.~Jeong and A.S.~Szalay, \emph{Toward accurate modelling of the
  non-linear matter bispectrum: standard perturbation theory and transients
  from initial conditions},
  \href{https://doi.org/10.1093/mnras/stv2525}{\emph{Monthly Notices of the
  Royal Astronomical Society} {\bfseries 455} (2015) 2945–2958}.

\bibitem{Sefusatti_2006}
E.~Sefusatti, M.~Crocce, S.~Pueblas and R.~Scoccimarro, \emph{Cosmology and the
  bispectrum}, \href{https://doi.org/10.1103/physrevd.74.023522}{\emph{Physical
  Review D} {\bfseries 74} (2006) }.

\bibitem{Song_2015}
Y.-S.~Song, A.~Taruya and A.~Oka, \emph{Cosmology with anisotropic galaxy
  clustering from the combination of power spectrum and bispectrum},
  \href{https://doi.org/10.1088/1475-7516/2015/08/007}{\emph{Journal of
  Cosmology and Astroparticle Physics} {\bfseries 2015} (2015) 007–007}.

\bibitem{Byun_2017}
J.~Byun, A.~Eggemeier, D.~Regan, D.~Seery and R.E.~Smith, \emph{Towards optimal
  cosmological parameter recovery from compressed bispectrum statistics},
  \href{https://doi.org/10.1093/mnras/stx1681}{\emph{Monthly Notices of the
  Royal Astronomical Society} {\bfseries 471} (2017) 1581–1618}.

\bibitem{Sefusatti_2012}
E.~Sefusatti, M.~Crocce and V.~Desjacques, \emph{The halo bispectrum in n-body
  simulations with non-gaussian initial conditions},
  \href{https://doi.org/10.1111/j.1365-2966.2012.21271.x}{\emph{Monthly Notices
  of the Royal Astronomical Society} {\bfseries 425} (2012) 2903–2930}.

\bibitem{Tellarini_2016}
M.~Tellarini, A.J.~Ross, G.~Tasinato and D.~Wands, \emph{Galaxy bispectrum,
  primordial non-gaussianity and redshift space distortions},
  \href{https://doi.org/10.1088/1475-7516/2016/06/014}{\emph{Journal of
  Cosmology and Astroparticle Physics} {\bfseries 2016} (2016) 014–014}.

\bibitem{karagiannis2020probing}
D.~Karagiannis, J.~Fonseca, R.~Maartens and S.~Camera, \emph{Probing primordial
  non-gaussianity with the bispectrum of future 21cm intensity maps},  2020.

\bibitem{dizgah2020primordial}
A.M.~Dizgah, M.~Biagetti, E.~Sefusatti, V.~Desjacques and J.~Noreña,
  \emph{Primordial non-gaussianity from biased tracers: Likelihood analysis of
  real-space power spectrum and bispectrum},  2020.

\bibitem{Chudaykin_2019}
A.~Chudaykin and M.M.~Ivanov, \emph{Measuring neutrino masses with large-scale
  structure: Euclid forecast with controlled theoretical error},
  \href{https://doi.org/10.1088/1475-7516/2019/11/034}{\emph{Journal of
  Cosmology and Astroparticle Physics} {\bfseries 2019} (2019) 034–034}.

\bibitem{Hahn_2020}
C.~Hahn, F.~Villaescusa-Navarro, E.~Castorina and R.~Scoccimarro,
  \emph{Constraining m$\nu$ with the bispectrum. part i. breaking parameter
  degeneracies},
  \href{https://doi.org/10.1088/1475-7516/2020/03/040}{\emph{Journal of
  Cosmology and Astroparticle Physics} {\bfseries 2020} (2020) 040–040}.

\bibitem{hahn2020constraining}
C.~Hahn and F.~Villaescusa-Navarro, \emph{Constraining $m_\nu$ with the
  bispectrum ii: The total information content of the galaxy bispectrum},
  2020.

\bibitem{kamalinejad2020nondegenerate}
F.~Kamalinejad and Z.~Slepian, \emph{A non-degenerate neutrino mass signature
  in the galaxy bispectrum},  2020.

\bibitem{Gil_Mar_n_2016}
H.~Gil-Marín, W.J.~Percival, L.~Verde, J.R.~Brownstein, C.-H.~Chuang,
  F.-S.~Kitaura et~al., \emph{The clustering of galaxies in the sdss-iii baryon
  oscillation spectroscopic survey: Rsd measurement from the power spectrum and
  bispectrum of the dr12 boss galaxies},
  \href{https://doi.org/10.1093/mnras/stw2679}{\emph{Monthly Notices of the
  Royal Astronomical Society} {\bfseries 465} (2016) 1757–1788}.

\bibitem{Slepian:2015hca}
Z.~Slepian et~al., \emph{{The large-scale three-point correlation function of
  the SDSS BOSS DR12 CMASS galaxies}},
  \href{https://doi.org/10.1093/mnras/stw3234}{\emph{Mon. Not. Roy. Astron.
  Soc.} {\bfseries 468} (2017) 1070}
  [\href{https://arxiv.org/abs/1512.02231}{{\ttfamily 1512.02231}}].

\bibitem{Slepian_2017}
Z.~Slepian, D.J.~Eisenstein, J.R.~Brownstein, C.-H.~Chuang, H.~Gil-Marín,
  S.~Ho et~al., \emph{Detection of baryon acoustic oscillation features in the
  large-scale three-point correlation function of sdss boss dr12 cmass
  galaxies}, \href{https://doi.org/10.1093/mnras/stx488}{\emph{Monthly Notices
  of the Royal Astronomical Society} {\bfseries 469} (2017) 1738–1751}.

\bibitem{Cabass/etal:2021}
G.~{Cabass}, M.M.~{Ivanov}, O.H.E.~{Philcox}, M.~{Simonovi{\'c}} and
  M.~{Zaldarriaga}, \emph{{Constraints on Single-Field Inflation from the BOSS
  Galaxy Survey}}, {\emph{arXiv e-prints} (2022) arXiv:2201.07238}
  [\href{https://arxiv.org/abs/2201.07238}{{\ttfamily 2201.07238}}].

\bibitem{Bernardeau_2002}
F.~Bernardeau, S.~Colombi, E.~Gaztañaga and R.~Scoccimarro, \emph{Large-scale
  structure of the universe and cosmological perturbation theory},
  \href{https://doi.org/10.1016/s0370-1573(02)00135-7}{\emph{Physics Reports}
  {\bfseries 367} (2002) 1–248}.

\bibitem{Jeong_2006}
D.~Jeong and E.~Komatsu, \emph{Perturbation theory reloaded: Analytical
  calculation of nonlinearity in baryonic oscillations in the real‐space
  matter power spectrum}, \href{https://doi.org/10.1086/507781}{\emph{The
  Astrophysical Journal} {\bfseries 651} (2006) 619–626}.

\bibitem{Nishimichi_2009}
T.~Nishimichi, A.~Shirata, A.~Taruya, K.~Yahata, S.~Saito, Y.~Suto et~al.,
  \emph{Modeling nonlinear evolution of baryon acoustic oscillations:
  Convergence regime of $n$-body simulations and analytic models},
  \href{https://doi.org/10.1093/pasj/61.2.321}{\emph{Publications of the
  Astronomical Society of Japan} {\bfseries 61} (2009) 321–332}.

\bibitem{Taruya_2009}
A.~Taruya, T.~Nishimichi, S.~Saito and T.~Hiramatsu, \emph{Nonlinear evolution
  of baryon acoustic oscillations from improved perturbation theory in real and
  redshift spaces},
  \href{https://doi.org/10.1103/physrevd.80.123503}{\emph{Physical Review D}
  {\bfseries 80} (2009) }.

\bibitem{Osato_2019}
K.~Osato, T.~Nishimichi, F.~Bernardeau and A.~Taruya, \emph{Perturbation theory
  challenge for cosmological parameters estimation: Matter power spectrum in
  real space}, \href{https://doi.org/10.1103/physrevd.99.063530}{\emph{Physical
  Review D} {\bfseries 99} (2019) }.

\bibitem{Steele_2021}
T.~Steele and T.~Baldauf, \emph{Precise calibration of the one-loop bispectrum
  in the effective field theory of large scale structure},
  \href{https://doi.org/10.1103/physrevd.103.023520}{\emph{Physical Review D}
  {\bfseries 103} (2021) }.

\bibitem{Lazanu_2018}
A.~Lazanu and M.~Liguori, \emph{The two and three-loop matter bispectrum in
  perturbation theories},
  \href{https://doi.org/10.1088/1475-7516/2018/04/055}{\emph{Journal of
  Cosmology and Astroparticle Physics} {\bfseries 2018} (2018) 055}.

\bibitem{Baldauf_2021}
T.~Baldauf, M.~Garny, P.~Taule and T.~Steele, \emph{Two-loop bispectrum of
  large-scale structure},
  \href{https://doi.org/10.1103/physrevd.104.123551}{\emph{Physical Review D}
  {\bfseries 104} (2021) }.

\bibitem{Villaescusa_Navarro_2020}
F.~Villaescusa-Navarro, C.~Hahn, E.~Massara, A.~Banerjee, A.M.~Delgado,
  D.K.~Ramanah et~al., \emph{The quijote simulations},
  \href{https://doi.org/10.3847/1538-4365/ab9d82}{\emph{The Astrophysical
  Journal Supplement Series} {\bfseries 250} (2020) 2}.

\bibitem{Planck2018}
{Planck Collaboration}, N.~{Aghanim}, Y.~{Akrami}, M.~{Ashdown}, J.~{Aumont},
  C.~{Baccigalupi} et~al., \emph{{Planck 2018 results. VI. Cosmological
  parameters}}, \href{https://doi.org/10.1051/0004-6361/201833910}{\emph{\aap}
  {\bfseries 641} (2020) A6}
  [\href{https://arxiv.org/abs/1807.06209}{{\ttfamily 1807.06209}}].

\bibitem{Hand_2018}
N.~Hand, Y.~Feng, F.~Beutler, Y.~Li, C.~Modi, U.~Seljak et~al., \emph{nbodykit:
  An open-source, massively parallel toolkit for large-scale structure},
  \href{https://doi.org/10.3847/1538-3881/aadae0}{\emph{The Astronomical
  Journal} {\bfseries 156} (2018) 160}.

\bibitem{Sefusatti_2016}
E.~Sefusatti, M.~Crocce, R.~Scoccimarro and H.M.P.~Couchman, \emph{Accurate
  estimators of correlation functions in fourier space},
  \href{https://doi.org/10.1093/mnras/stw1229}{\emph{Monthly Notices of the
  Royal Astronomical Society} {\bfseries 460} (2016) 3624–3636}.

\bibitem{Scoccimarro_2015}
R.~Scoccimarro, \emph{Fast estimators for redshift-space clustering},
  \href{https://doi.org/10.1103/physrevd.92.083532}{\emph{Physical Review D}
  {\bfseries 92} (2015) }.

\bibitem{Tomlinson_2019}
J.~Tomlinson, D.~Jeong and J.~Kim, \emph{Efficient parallel algorithm for
  estimating higher-order polyspectra},
  \href{https://doi.org/10.3847/1538-3881/ab3223}{\emph{The Astronomical
  Journal} {\bfseries 158} (2019) 116}.

\bibitem{WMAP2013}
C.L.~{Bennett}, D.~{Larson}, J.L.~{Weiland}, N.~{Jarosik}, G.~{Hinshaw},
  N.~{Odegard} et~al., \emph{{Nine-year Wilkinson Microwave Anisotropy Probe
  (WMAP) Observations: Final Maps and Results}},
  \href{https://doi.org/10.1088/0067-0049/208/2/20}{\emph{\apjs} {\bfseries
  208} (2013) 20} [\href{https://arxiv.org/abs/1212.5225}{{\ttfamily
  1212.5225}}].

\bibitem{Planckfnl2020}
{Planck Collaboration}, Y.~{Akrami}, F.~{Arroja}, M.~{Ashdown}, J.~{Aumont},
  C.~{Baccigalupi} et~al., \emph{{Planck 2018 results. IX. Constraints on
  primordial non-Gaussianity}},
  \href{https://doi.org/10.1051/0004-6361/201935891}{\emph{\aap} {\bfseries
  641} (2020) A9} [\href{https://arxiv.org/abs/1905.05697}{{\ttfamily
  1905.05697}}].

\bibitem{IASMatrixFFT2018}
M.~{Simonovi{\'c}}, T.~{Baldauf}, M.~{Zaldarriaga}, J.J.~{Carrasco} and
  J.A.~{Kollmeier}, \emph{{Cosmological perturbation theory using the FFTLog:
  formalism and connection to QFT loop integrals}},
  \href{https://doi.org/10.1088/1475-7516/2018/04/030}{\emph{\jcap} {\bfseries
  2018} (2018) 030} [\href{https://arxiv.org/abs/1708.08130}{{\ttfamily
  1708.08130}}].

\bibitem{Lewis_2000}
A.~Lewis, A.~Challinor and A.~Lasenby, \emph{Efficient computation of cosmic
  microwave background anisotropies in closed friedmann‐robertson‐walker
  models}, \href{https://doi.org/10.1086/309179}{\emph{The Astrophysical
  Journal} {\bfseries 538} (2000) 473–476}.

\bibitem{Taruya_2018}
A.~Taruya, T.~Nishimichi and D.~Jeong, \emph{Grid-based calculation for
  perturbation theory of large-scale structure},
  \href{https://doi.org/10.1103/physrevd.98.103532}{\emph{Physical Review D}
  {\bfseries 98} (2018) }.

\bibitem{Slosar_2008}
A.~Slosar, C.~Hirata, U.~Seljak, S.~Ho and N.~Padmanabhan, \emph{Constraints on
  local primordial non-gaussianity from large scale structure},
  \href{https://doi.org/10.1088/1475-7516/2008/08/031}{\emph{Journal of
  Cosmology and Astroparticle Physics} {\bfseries 2008} (2008) 031}.

\bibitem{Hamaus2011}
N.~{Hamaus}, U.~{Seljak} and V.~{Desjacques}, \emph{{Optimal constraints on
  local primordial non-Gaussianity from the two-point statistics of large-scale
  structure}}, \href{https://doi.org/10.1103/PhysRevD.84.083509}{\emph{\prd}
  {\bfseries 84} (2011) 083509}
  [\href{https://arxiv.org/abs/1104.2321}{{\ttfamily 1104.2321}}].

\bibitem{Giannantonio2014}
T.~Giannantonio, A.J.~Ross, W.J.~Percival, R.~Crittenden, D.~Bacher,
  M.~Kilbinger et~al., \emph{Improved primordial non-gaussianity constraints
  from measurements of galaxy clustering and the integrated sachs-wolfe
  effect}, \href{https://doi.org/10.1103/PhysRevD.89.023511}{\emph{Phys. Rev.
  D} {\bfseries 89} (2014) 023511}.

\bibitem{Agarwal_2014}
N.~Agarwal, S.~Ho and S.~Shandera, \emph{Constraining the initial conditions of
  the universe using large scale structure},
  \href{https://doi.org/10.1088/1475-7516/2014/02/038}{\emph{Journal of
  Cosmology and Astroparticle Physics} {\bfseries 2014} (2014) 038}.

\bibitem{Barreira2020}
A.~{Barreira}, G.~{Cabass}, F.~{Schmidt}, A.~{Pillepich} and D.~{Nelson},
  \emph{{Galaxy bias and primordial non-Gaussianity: insights from galaxy
  formation simulations with IllustrisTNG}},
  \href{https://doi.org/10.1088/1475-7516/2020/12/013}{\emph{\jcap} {\bfseries
  2020} (2020) 013} [\href{https://arxiv.org/abs/2006.09368}{{\ttfamily
  2006.09368}}].

\bibitem{Morad2021}
A.~{Moradinezhad Dizgah}, M.~{Biagetti}, E.~{Sefusatti}, V.~{Desjacques} and
  J.~{Nore{\~n}a}, \emph{{Primordial non-Gaussianity from biased tracers:
  likelihood analysis of real-space power spectrum and bispectrum}},
  \href{https://doi.org/10.1088/1475-7516/2021/05/015}{\emph{\jcap} {\bfseries
  2021} (2021) 015} [\href{https://arxiv.org/abs/2010.14523}{{\ttfamily
  2010.14523}}].

\bibitem{Rezaie_2021}
M.~Rezaie, A.J.~Ross, H.-J.~Seo, E.-M.~Mueller, W.J.~Percival, G.~Merz et~al.,
  \emph{Primordial non-gaussianity from the completed {SDSS}-{IV} extended
  baryon oscillation spectroscopic survey {\textendash} i: Catalogue
  preparation and systematic mitigation},
  \href{https://doi.org/10.1093/mnras/stab1730}{\emph{Monthly Notices of the
  Royal Astronomical Society} {\bfseries 506} (2021) 3439}.

\bibitem{Sefusatti_2010}
E.~Sefusatti, M.~Crocce and V.~Desjacques, \emph{The matter bispectrum in
  n-body simulations with non-gaussian initial conditions},
  \href{https://doi.org/10.1111/j.1365-2966.2010.16723.x}{\emph{Monthly Notices
  of the Royal Astronomical Society} (2010) no–no}.

\bibitem{Yankelevich2018}
V.~Yankelevich and C.~Porciani, \emph{Cosmological information in the
  redshift-space bispectrum},
  \href{https://doi.org/10.1093/mnras/sty3143}{\emph{Monthly Notices of the
  Royal Astronomical Society} {\bfseries 483} (2018) 2078–2099}.

\bibitem{Oddo_2020}
A.~Oddo, E.~Sefusatti, C.~Porciani, P.~Monaco and A.G.~Sánchez, \emph{Toward a
  robust inference method for the galaxy bispectrum: likelihood function and
  model selection},
  \href{https://doi.org/10.1088/1475-7516/2020/03/056}{\emph{Journal of
  Cosmology and Astroparticle Physics} {\bfseries 2020} (2020) 056–056}.

\bibitem{Jeong/Komatsu:2009}
D.~{Jeong} and E.~{Komatsu}, \emph{{Primordial Non-Gaussianity, Scale-dependent
  Bias, and the Bispectrum of Galaxies}},
  \href{https://doi.org/10.1088/0004-637X/703/2/1230}{\emph{\apj} {\bfseries
  703} (2009) 1230} [\href{https://arxiv.org/abs/0904.0497}{{\ttfamily
  0904.0497}}].

\bibitem{Eisenstein_1998}
D.J.~Eisenstein and W.~Hu, \emph{Baryonic features in the matter transfer
  function}, \href{https://doi.org/10.1086/305424}{\emph{The Astrophysical
  Journal} {\bfseries 496} (1998) 605–614}.

\bibitem{Jeong/Komatsu:2006}
D.~{Jeong} and E.~{Komatsu}, \emph{{Perturbation Theory Reloaded: Analytical
  Calculation of Nonlinearity in Baryonic Oscillations in the Real-Space Matter
  Power Spectrum}}, \href{https://doi.org/10.1086/507781}{\emph{\apj}
  {\bfseries 651} (2006) 619}
  [\href{https://arxiv.org/abs/astro-ph/0604075}{{\ttfamily
  astro-ph/0604075}}].

\bibitem{WMAP32007}
D.N.~{Spergel}, R.~{Bean}, O.~{Dor{\'e}}, M.R.~{Nolta}, C.L.~{Bennett},
  J.~{Dunkley} et~al., \emph{{Three-Year Wilkinson Microwave Anisotropy Probe
  (WMAP) Observations: Implications for Cosmology}},
  \href{https://doi.org/10.1086/513700}{\emph{\apjs} {\bfseries 170} (2007)
  377} [\href{https://arxiv.org/abs/astro-ph/0603449}{{\ttfamily
  astro-ph/0603449}}].

\bibitem{Gualdi_2021}
D.~Gualdi, S.~Novell, H.~Gil-Mar{\'{\i}}n and L.~Verde, \emph{Matter
  trispectrum: theoretical modelling and comparison to n-body simulations},
  \href{https://doi.org/10.1088/1475-7516/2021/01/015}{\emph{Journal of
  Cosmology and Astroparticle Physics} {\bfseries 2021} (2021) 015}.

\bibitem{Gualdi_2021b}
D.~Gualdi, H.~Gil-Mar{\'{\i}}n and L.~Verde, \emph{Joint analysis of
  anisotropic power spectrum, bispectrum and trispectrum: application to n-body
  simulations},
  \href{https://doi.org/10.1088/1475-7516/2021/07/008}{\emph{Journal of
  Cosmology and Astroparticle Physics} {\bfseries 2021} (2021) 008}.

\bibitem{Cahn2021}
R.N.~{Cahn}, Z.~{Slepian} and J.~{Hou}, \emph{{A Test for Cosmological Parity
  Violation Using the 3D Distribution of Galaxies}}, {\emph{arXiv e-prints}
  (2021) arXiv:2110.12004} [\href{https://arxiv.org/abs/2110.12004}{{\ttfamily
  2110.12004}}].

\bibitem{Philcox2021}
O.H.E.~{Philcox}, J.~{Hou} and Z.~{Slepian}, \emph{{A First Detection of the
  Connected 4-Point Correlation Function of Galaxies Using the BOSS CMASS
  Sample}}, {\emph{arXiv e-prints} (2021) arXiv:2108.01670}
  [\href{https://arxiv.org/abs/2108.01670}{{\ttfamily 2108.01670}}].

\bibitem{Gualdi2022}
D.~{Gualdi} and L.~{Verde}, \emph{{Integrated trispectrum detection from BOSS
  DR12 NGC CMASS}}, {\emph{arXiv e-prints} (2022) arXiv:2201.06932}
  [\href{https://arxiv.org/abs/2201.06932}{{\ttfamily 2201.06932}}].

\bibitem{Baumann2012}
D.~{Baumann}, A.~{Nicolis}, L.~{Senatore} and M.~{Zaldarriaga},
  \emph{{Cosmological non-linearities as an effective fluid}},
  \href{https://doi.org/10.1088/1475-7516/2012/07/051}{\emph{\jcap} {\bfseries
  2012} (2012) 051} [\href{https://arxiv.org/abs/1004.2488}{{\ttfamily
  1004.2488}}].

\bibitem{Carrasco_2012}
J.J.M.~Carrasco, M.P.~Hertzberg and L.~Senatore, \emph{The effective field
  theory of cosmological large scale structures},
  \href{https://doi.org/10.1007/jhep09(2012)082}{\emph{Journal of High Energy
  Physics} {\bfseries 2012} (2012) }.

\bibitem{Carrasco_2014}
J.J.M.~Carrasco, S.~Foreman, D.~Green and L.~Senatore, \emph{The 2-loop matter
  power spectrum and the {IR}-safe integrand},
  \href{https://doi.org/10.1088/1475-7516/2014/07/056}{\emph{Journal of
  Cosmology and Astroparticle Physics} {\bfseries 2014} (2014) 056}.

\bibitem{Carrasco_2014b}
J.J.M.~Carrasco, S.~Foreman, D.~Green and L.~Senatore, \emph{The effective
  field theory of large scale structures at two loops},
  \href{https://doi.org/10.1088/1475-7516/2014/07/057}{\emph{Journal of
  Cosmology and Astroparticle Physics} {\bfseries 2014} (2014) 057}.

\bibitem{Hertzberg2014}
M.P.~Hertzberg, \emph{Effective field theory of dark matter and structure
  formation: Semianalytical results},
  \href{https://doi.org/10.1103/PhysRevD.89.043521}{\emph{Phys. Rev. D}
  {\bfseries 89} (2014) 043521}.

\bibitem{Senatore_2015}
L.~Senatore and M.~Zaldarriaga, \emph{The {IR}-resummed effective field theory
  of large scale structures},
  \href{https://doi.org/10.1088/1475-7516/2015/02/013}{\emph{Journal of
  Cosmology and Astroparticle Physics} {\bfseries 2015} (2015) 013}.

\bibitem{Taruya2012}
A.~Taruya, F.~Bernardeau, T.~Nishimichi and S.~Codis, \emph{Direct and fast
  calculation of regularized cosmological power spectrum at two-loop order},
  \href{https://doi.org/10.1103/PhysRevD.86.103528}{\emph{Phys. Rev. D}
  {\bfseries 86} (2012) 103528}.

\bibitem{Matsubara2008}
T.~Matsubara, \emph{Resumming cosmological perturbations via the lagrangian
  picture: One-loop results in real space and in redshift space},
  \href{https://doi.org/10.1103/PhysRevD.77.063530}{\emph{Phys. Rev. D}
  {\bfseries 77} (2008) 063530}.

\bibitem{Eggemeier/etal:2020}
A.~{Eggemeier}, R.~{Scoccimarro}, M.~{Crocce}, A.~{Pezzotta} and
  A.G.~{S{\'a}nchez}, \emph{{Testing one-loop galaxy bias: Power spectrum}},
  \href{https://doi.org/10.1103/PhysRevD.102.103530}{\emph{\prd} {\bfseries
  102} (2020) 103530} [\href{https://arxiv.org/abs/2006.09729}{{\ttfamily
  2006.09729}}].

\bibitem{Alkhanishvili_2022}
D.~Alkhanishvili, C.~Porciani, E.~Sefusatti, M.~Biagetti, A.~Lazanu, A.~Oddo
  et~al., \emph{The reach of next-to-leading-order perturbation theory for the
  matter bispectrum},
  \href{https://doi.org/10.1093/mnras/stac567}{\emph{Monthly Notices of the
  Royal Astronomical Society} (2022) }.

\end{thebibliography}\endgroup

\end{document}